\documentclass[reprint, showpacs, groupedaddress
 amsmath,amssymb,
 aps,
 natbib,
]{revtex4-1}

\usepackage{graphicx}
\usepackage{dcolumn}
\usepackage{bm}
\usepackage{hyperref}
\usepackage{amssymb}
\usepackage{mathrsfs}
\usepackage{amsmath,fleqn}

\newcommand{\bea}{\begin{eqnarray}}
\newcommand{\eea}{\end{eqnarray}}
\newcommand{\ba}{\begin{array}}
\newcommand{\ea}{\end{array}}

\newcommand{\be}{\begin{equation}}

\newcommand{\ee}{\end{equation}}
\newcommand{\bt}{\begin{teo}}
\newcommand{\et}{\end{teo}}

\newcommand{\om}{\omega}

\newcommand{\blo}{\beta_{\rm loc}}

\newcommand{\La}{\Lambda}
\newcommand{\linf}{{\cal L}}

\begin{document}

\preprint{APS/123-QED}

\title{Statistical properties of the localization measure in a finite-dimensional model of the quantum kicked rotator }

\author{Thanos Manos}
\email{thanos.manos@uni-mb.si}
\affiliation{CAMTP - Center for Applied Mathematics and Theoretical Physics, University of Maribor, Krekova 2, SI-2000 Maribor, Slovenia}
\affiliation{School of Applied Sciences, University of Nova Gorica, Vipavska 11c, SI-5270 Ajdov\v s\v cina, Slovenia}
\affiliation{Institute of Neuroscience and Medicine Neuromodulation
(INM-7), Research Center J\"ulich, D-52425 J\"ulich, Germany}

\author{Marko Robnik}
\email{Robnik@uni-mb.si}
\affiliation{%
 CAMTP - Center for Applied Mathematics and Theoretical Physics, University of Maribor, Krekova 2, SI-2000 Maribor, Slovenia}

\date{\today}

\begin{abstract}
We study the quantum kicked rotator in the classically fully chaotic regime $K=10$ and for various values of the quantum parameter $k$ using Izrailev's $N$-dimensional model for various $N \le 3000$, which in the limit $N \rightarrow \infty$ tends to the exact quantized kicked rotator. By numerically calculating the eigenfunctions in the basis of the angular momentum we find that the localization length ${\cal L}$ for fixed parameter values has a certain distribution, in fact its inverse is Gaussian distributed, in analogy and in connection with the distribution of finite time Lyapunov exponents of Hamilton systems. However, unlike the case of the finite time Lyapunov exponents, this distribution is found to be independent of $N$, and thus survives the limit $N=\infty$. This is different from the tight-binding model of Anderson localization. The reason is that the finite bandwidth approximation of the underlying Hamilton dynamical system in the Shepelyansky picture (D.L. Shepelyansky, {\em Phys. Rev. Lett.} {\bf 56}, 677 (1986)) does not apply rigorously. This observation explains the strong fluctuations in the scaling laws of the kicked rotator, such as e.g. the entropy localization measure as a function of the scaling parameter $\Lambda={\cal L}/N$, where $\linf$ is the theoretical value of the localization length in the semiclassical approximation. These results call for a more refined theory of the localization length in the quantum kicked rotator and in similar Floquet systems, where we must predict not only the mean value of the inverse of the localization length $\linf$ but also its (Gaussian) distribution, in particular the variance. In order to complete our studies we numerically analyze the related behavior of finite time Lyapunov exponents in the standard map and of the 2$\times2$ transfer matrix formalism. This paper is extending our recent work (T. Manos and M. Robnik, {\em Phys. Rev. E} {\bf 87}, 062905 (2013)).


\end{abstract}

\maketitle


\section{Introduction \label{intro}}

Time-periodic (Floquet) quantum systems, whose classical analog is fully chaotic and diffusive, typically exhibit dynamical localization \cite{Stoe,Haake}, if a certain semiclassical condition is satisfied,
as explained below. We study the periodically kicked rotator in the classically fully chaotic regime $K=10$ using Izrailev's $N$-dimensional model
\cite{Izr1986,Izr1987,Izr1989,Izr1990} for various $N \le 3000$, which in the limit $N \rightarrow \infty$  tends to the quantized kicked rotator. We restrict our analysis to the case $K=10$ because this is empirically
the most typical uniformly chaotic regime, apparently free of any islands of stability or acceleration modes \cite{ManRob2014}. Due to the finiteness of $N$ the observed (dimensionless) localization length of the eigenfunctions in the space of the angular momentum quantum number does not possess a sharply defined value, but has a certain distribution instead.  Its reciprocal value is almost Gaussian distributed. This might be expected on the analogy with the finite time Lyapunov exponents in the Hamiltonian dynamical systems. In order to corroborate the theoretical findings on this topics we perform in Secs.~\ref{sec4} and \ref{sec6} the numerical analysis of the finite time Lyapunov exponents in the standard map (classical kicked rotator), especially the decay of the variance. Indeed, in the Shepelyansky picture \cite{She1986} the localization length can be obtained as the inverse of the smallest positive Lyapunov exponent of a finite $2k$-dimensional Hamilton system associated with the band matrix representation of the quantum kicked rotator, where $k$ is the quantum kick parameter (to be precisely defined below).  In this picture, $N$ plays the role of time. However, unlike the chaotic classical maps or products of transfer matrices in the Anderson tight-binding approximation, where the mean value of the finite time Lyapunov exponents is usually equal to their asymptotical value of infinite time and the variance decreases inversely with time, as we also carefully checked (see Secs.~\ref{sec5} and \ref{sec6}), here the distribution is found to be independent of $N$: It has a nonzero variance even in the limit $N\rightarrow \infty$. The reason is that the quantum kicked rotator at $N=\infty$ cannot be {\em exactly} modeled with finite bandwidth (equal to $2k$) band matrices, but only approximately, such that the underlying Hamilton system of the Shepelyansky picture has a growing dimension with $N$, implying asymptotically an infinite set of Lyapunov exponents and behavior different from the finite dimensional Hamiltonian systems. The observation of the distribution of the localization length around its mean value with finite variance also explains the strong fluctuations in the scaling laws of the kicked rotator, such as e.g. the entropy localization measure as a function of the theoretical scaling parameter $\Lambda$, to be discussed below. On the other hand, the two different empirical localization measures, namely the mean localization length as extracted directly from the exponentially localized eigenfunctions and the measure based on the information entropy of the eigenstates, are perfectly well linearly connected and thus equivalent. Therefore these results call for a refined theory of the localization length in the quantum kicked rotator and similar systems, where we must predict not only the mean value of the inverse localization length but also its (Gaussian) distribution, in particular the variance. This paper is a follow-up paper of our recent work \cite{ManRob2013} (Manos and Robnik 2013).

The time-independent and time-periodic systems have much in common when
discussing the localization properties of the chaotic eigenstates and of the
corresponding energy spectra. The main result of stationary quantum chaos (or wave chaos) \cite{Stoe,Haake,Rob1998} is the discovery that in classically fully chaotic, ergodic, autonomous Hamilton systems with the purely discrete spectrum the fluctuations of the energy spectrum around its mean behavior obey the statistical laws described by the Gaussian Random Matrix Theory (RMT) \cite{Mehta,GMW}, provided that we are in the sufficiently deep semiclassical limit. The latter semiclassical condition means that all relevant classical transport times are smaller than the so-called Heisenberg time, or break time, given by $t_H=2\pi\hbar/\Delta E$, where $h=2\pi\hbar$ is the Planck constant and $\Delta E$ is the mean energy level spacing, such that the mean energy level density is $\rho(E) =1/\Delta E$. This statement is known as the Bohigas - Giannoni - Schmit (BGS) conjecture and goes back to their pioneering paper in 1984 \cite{BGS}, although some preliminary ideas were published in \cite{Cas}.  Since $\Delta E \propto \hbar^f$, where $f$ is the number of degrees of freedom (= the dimension of the configuration space), we see that for sufficiently small $\hbar$ the stated condition will always be satisfied. Alternatively, fixing the $\hbar$, we can go to high energies such that the classical transport times become smaller than $t_H$. The role of the antiunitary symmetries that classify the statistics in terms of GOE, GUE or GSE (ensembles of RMT) has been elucidated in \cite{RB1986}, see also \cite{Rob1986}, and \cite{Stoe,Haake,Rob1998,Mehta}. The theoretical foundation for the BGS conjecture has been initiated first by Berry \cite{Berry1985}, and later further developed by Richter and Sieber \cite{Sieber}, arriving finally in the almost-final proof proposed by the group of F. Haake \cite{Mueller1,Mueller2,Mueller3,Mueller4}.

Here it must be emphasized again that considering the chaotic eigenstates
and their dynamical localization properties 
there are strong analogies between the time-periodic systems (like the kicked
rotator) and time-independent systems (like static billiards) \cite{Pro2000}, where the Brody distribution \cite{Bro1973,Bro1981} plays a key role, as discussed in  \cite{BatRob2010, BMR2013,BatRob2013,BatRob2013A,ManRob2013}. 

The paper is organized as follows: In Sec.~\ref{sec2} we define the model, in Sec.~\ref{sec3} we present the evidence for and the description of the distribution of the localization measures, in Sec.~\ref{sec4} we study the finite time Lyapunov exponents of the classical standard mapping as a generic example of a chaotic area preserving mapping, in Sec.~\ref{sec5} we study the finite time Lyapunov exponents of the product of two-dimensional random symplectic matrices describing the tight-binding model of Anderson localization, in Sec.~\ref{sec6} we present the high precision numerical results about the decay of the variance of the distribution of the
finite Lyapunov exponents of Sec.~\ref{sec4} and \ref{sec5}, and in Sec.~\ref{sec7} we conclude and discuss the results in the broader theoretical perspective.

\section{The  kicked rotator,  the Izrailev model  and  the dynamical localization }
\label{sec2}

The kicked rotator was introduced by Casati, Chirikov, Ford and Izrailev in 1979 \cite{CCFI79}. Here we follow our notation \cite{ManRob2013}. The Hamiltonian function is
\be  \label{KR}
H= \frac{p^2}{2I} + V_0 \,\delta_T(t)\,\cos \theta.
\ee
Here $p$ is the (angular) momentum, $I$ the moment of inertia, $V_0$ is the strength of the periodic kicking, $\theta\in [0,2\pi)$ is the (canonically conjugate, rotation) angle, and $\delta_T(t)$ is the periodic
Dirac delta function with period $T$. Between the kicks the rotation
is free, and thus the dynamics can be reduced to the standard mapping,

\be \label{SM1}
p_{n+1} = p_n + V_0 \sin \theta_{n+1},\;\;\; \theta_{n+1} = \theta_n + \frac{T}{I} p_n,
\ee
as introduced in \cite{T69,F72,C79}. The quantities $(\theta_n, p_n)$ refer to their  values just immediately after the $n$-th kick.
By using new dimensionless momentum $P_n = p_nT/I$, we get
\be \label{SM2}
P_{n+1} = P_n + K \sin \theta_{n+1},\;\;\; \theta_{n+1} = \theta_n  + P_n,
\ee
where the system has now a single classical {\em dimensionless}
control parameter $K=V_0 T/I$.

The quantum kicked rotator (QKR) is the quantized version of Eq.~(\ref{KR}),
namely
\be \label{QKR}
\hat{H} =-\frac{\hbar^2}{2I} \frac{\partial^2}{\partial\theta^2} +
V_0\, \delta_T(t)\,\cos \theta .
\ee
The Floquet operator $\hat{F}$ acting on the wavefunctions (probability amplitudes) $\psi(\theta)$, $\theta \in[0,2\pi)$, upon each period (of length $T$) can be written as (see e.g. \cite{Stoe}, Chapter 4)
\be \label{Fop}
\hat{F} = \exp \left( -\frac{iV_0}{\hbar} \cos\theta\right)
\exp\left(-\frac{i\hbar T}{2I}\frac{\partial^2}{\partial\theta^2}\right),
\ee
where now we have two {\em dimensionless} quantum control parameters

\be \label{qpar}
k=\frac{V_0}{\hbar}, \;\;\; \tau= \frac{\hbar T}{I},
\ee
which satisfy the relationship $K = k\tau = V_0 T/I$, $K$ being the classical {\em dimensionless} control parameter of Eq.~(\ref{SM2}). By using the angular momentum eigenfunctions

\be \label{Eigenf}
\langle\theta|n\rangle = a_n(\theta) = \frac{1}{\sqrt{2\pi}} \exp (i\,n\,\theta),
\ee
where $n$ is any integer, we find the matrix elements of $\hat{F}$, namely

\begin{flalign} \label{Fmatrix}
F_{m\,n} = \langle m|\hat{F}|n\rangle = \exp\left(-\frac{i\tau}{2} n^2\right) i^{n-m} J_{n-m} (k),
\end{flalign}
where $J_{\nu}(k)$ is the $\nu$-th order Bessel function. For a wavefunction
$\psi (\theta)$ we shall denote its angular momentum component (Fourier
component) by

\begin{flalign} \label{Fourier} \nonumber
u_n = \langle n|\psi\rangle = \int_0^{2\pi} a_n^{*}(\theta) \psi(\theta)\,d\theta= \\ =\frac{1}{\sqrt{2\pi}} \int_0^{2\pi} \psi(\theta) \exp(-in\theta)\,d\theta.
\end{flalign}
The QKR has very complex dynamics and spectral properties. As the phase space is infinite (cylinder), $p\in (-\infty, +\infty), \theta\in[0,2\pi)$, the spectrum of the eigenphases of $\hat{F}$, denoted by $\phi_n$, or the associated quasienergies $\hbar\om_n= \hbar \phi_n/T$, introduced by Zeldovich  \cite{Zel1966}, can be continuous, or discrete \cite{IS1979a,IS1979b,IS1980a,IS1980b}.

The asymptotic localized eigenstates are {\em exponentially localized}. The (dimensionless) theoretical localization length in the space of the angular momentum quantum numbers is given below, and is equal (after introducing some numerical correction factor $\alpha_{\mu}$) to the dimensionless localization time $t_{\rm loc}$  [Eq.~(\ref{finallinf}), given below]. We denote it unlike in reference \cite{Izr1990} and \cite{ManRob2013} by $\linf$.  Therefore, an exponentially localized eigenfunction centered at $m$ in the angular momentum space [Eq.~(\ref{Eigenf})] has the following form
\be \label{exploc} |u_n|^2 \approx \frac{1}{\linf} \exp\left(-\frac{2|m-n|}{\linf}\right),
\ee
where $u_n$ is the probability amplitude [Eq.~(\ref{Fourier})] of the localized wavefunction $\psi(\theta)$. The argument leading to $t_{\rm loc}$ in Eq.~(\ref{finallinf}) given below originates from the observation of the dynamical localization by Casati et al \cite{CCFI79}, and in particular from \cite{CIS1981}, and is well explained in \cite{Stoe}, in case of normal diffusion, whilst for general anomalous diffusion we gave a theoretical argument in \cite{ManRob2013}. We shall denote $\sigma = 2/\linf$, and will
later on determine the $\sigma$'s directly from the individual numerically calculated eigenstate.

The question arises, where do we see the  phenomena (spectral statistics, namely
Brody-like level spacing distribution) analogous  in the quantum chaos of time-independent bound systems with discrete spectrum?  To see these effects the system must have effectively finite dimension, because in the
infinite dimensional case we simply observe Poissonian statistics. Truncation of the infinite matrix $F_{mn}$ in Eq.~(\ref{Fmatrix}) in {\em tour de force} is not acceptable, even in the technical case of numerical computations, since after truncation the Floquet operator is no longer unitary.

The only way to obtain a quantum system which shall in this sense correspond to the classical dynamical system [Eqs.~(\ref{KR}), (\ref{SM1}) and (\ref{SM2})] is to introduce a finite $N$-dimensional matrix, which is symmetric unitary, and which in the limit $N\rightarrow\infty$ becomes the infinite dimensional system with the Floquet operator [Eq.~(\ref{Fop})]. The semiclassical limit is $k\rightarrow \infty$ and $\tau\rightarrow 0$, such that $K=k\tau ={\rm constant}$. As it is well known \cite{Izr1990}, for the reasons discussed above, the system behaves very similarly for rational and irrational values of $\tau/(4\pi)$. Such a $N$-dimensional model \cite{Izr1988} will be introduced below.

The generalized diffusion process of the standard map (\ref{SM2}) is defined by
\be \label{varp}
\langle(\Delta P)^2\rangle = D_{\mu}(K) n^{\mu},
\ee
where $n$ is the number of iterations (kicks), and the exponent $\mu$ is in the interval $[0,2)$, and all variables $P$, $\theta$ and $K$ are dimensionless. Here $D_{\mu}(K)$ is the {\bf generalized classical diffusion constant}. The averaging $\langle .\rangle$ is over an ensemble of initial conditions with fixed
$P$, specifically in our case $P=0$.
In case $\mu=1$ we have the normal diffusion, and $D_1(K)$ is then the normal diffusion constant, whilst in case of anomalous diffusion we observe subdiffusion when $0 < \mu < 1$,  or superdiffusion if $1 <\mu \le2$. In case $\mu=2$ we have the ballistic transport which is associated with the presence of accelerator modes (see below).

Following \cite{ManRob2013} we find that the dimensionless Heisenberg
time, also called break time or localization time, denoted by $t_{\rm loc}$,
in units of kicking period $T$, is equal to the dimensionless
localization length $\linf$

\be \label{finallinf}
\linf \approx t_{\rm loc} = \left( \alpha_{\mu} \frac{D_{\mu}(K)}{\tau^2} \right)^{\frac{1}{2-\mu}}.
\ee
where $\alpha_{\mu}$ is a numerical constant to be determined empirically,
and in case of normal diffusion $\mu=1$ is close to $1/2$.

In case of the normal diffusion $\mu=1$, considered in the present
paper, the theoretical value of $D_1(K)$ is given in the literature, e.g. in \cite{Izr1990} or \cite{LL1992},

\begin{flalign} \label{Dcl}
D_{1}(K)=
\begin{cases}
 \frac{1}{2} K^2\left [1- 2J_2(K) \left (1-J_2(K) \right ) \right ], \text{if} \ K \ge 4.5 \\
 0.15(K-K_{\rm cr})^3, \text{if} \ K_{cr} < K \le 4.5
\end{cases},
\end{flalign}
where $K_{\rm cr} \simeq 0.9716$ and $J_2(K)$ is the Bessel function. Here we neglect higher terms of order $K^{-2}$. In the present paper we shall consider
exclusively the case $K=10$,  which has been carefully checked to be
fully chaotic, without any regular islands, and well described by the
normal diffusion $\mu=1$, so that the above formula applies very well
\cite{ManRob2014}.

The motion of the QKR [Eq.~(\ref{QKR})] after one period $T$ of the $\psi$ wavefunction can be described also by the following symmetrized Floquet mapping, describing the evolution of the kicked rotator from the middle of a free rotation over a kick to the middle of the next free rotation, as follows
\begin{flalign} \label{Uoper}
& \psi(\theta,t+T) = \hat{U}\psi(\theta,t), \\ \nonumber
& \hat{U} = \exp \left ( i \frac{T\hbar}{4I}\frac{\partial^2}{\partial \theta^2} \right )\exp \left (-i\frac{V_0}{\hbar} \cos \theta \right)\exp \left (i \frac{T\hbar}{4I}\frac{\partial^2}{\partial \theta^2} \right).
\end{flalign}
Thus, the $\psi(\theta,t)$ function is determined in the middle of the rotation, between two successive kicks. The evolution operator $\hat{U}$ of the system corresponds to one period. 

 In the case $K\equiv k \tau \gg 1$ the motion is well known to be strongly chaotic, for $K=10$ certainly without any regular islands of stability, and also there are no accelerator modes, so that the diffusion is normal
($\mu=1$). We have carefully checked that the case $K=10$ is the closest
to the normal diffusion $\mu=1$  for all $K\in [0,70]$. The transition to classical mechanics is described by the limit $k \rightarrow \infty$, $\tau \rightarrow 0$ while $K=\rm{const}$. We shall consider the regimes on the
interval $3\le k\le 20$, but will concentrate mostly on the semiclassical regime $k\ge K$, where $\tau \le 1$.

In order to study how the localization affects the statistical properties of the quasienergy spectra, we use the model's representation in the momentum space with a finite number $N$ of levels \cite{Izr1988,Izr1990,Izr1986,Izr1987,Izr1989}, which we refer to as Izrailev model
\begin{flalign} \label{u_repres}
u_n(t+T) = \sum_{m=1}^{N} U_{nm}u_m(t), \ n,m=1,2,...,N \enspace .
\end{flalign}
The finite symmetric unitary matrix $U_{nm}$ determines the evolution of an $N$-dimensional vector, namely the Fourier transform $u_n(t)$ of $\psi(\theta,t)$, and is composed in the following way
\be \label{Unm}
    U_{nm}=\sum_{n'm'}G_{nm'}B_{n'm'}G_{n'm},
\ee
where $G_{ll'}=\exp \left (i\tau l^2/4 \right )\delta_{ll'}$ is a diagonal matrix corresponding to free rotation during a half period $T/2$, and  the matrix $B_{n'm'}$ describing the one kick has the following form
\begin{flalign} \label{Bnmoper}\nonumber
  & B_{n'm'}= \frac{1}{2N+1}\times \\ \nonumber
  & \sum_{l=1}^{2N+1} \left \{ \cos \left [ \left (n'-m' \right ) \frac{2 \pi l}{2N+1}\right ] - \cos \left [(n'+m')\frac{2 \pi l}{2N+1} \right ] \right \} \\
  & \times  \exp \left [-ik\cos \left (\frac{2 \pi l}{2N+1}\right ) \right ].
\end{flalign}
The Izrailev model in Eqs.~(\ref{u_repres}-\ref{Bnmoper}) with a finite number of states is considered as the quantum analogue of the classical standard mapping on the torus with closed momentum $p$ and phase $\theta$, where $U_{nm}$ describes only the odd states of the systems, i.e. $\psi(\theta)=-\psi(-\theta)$, provided we have the case of the quantum resonance, namely $\tau =4\pi r/(2N+1)$, where $r$ is a positive integer. The matrix (\ref{Bnmoper}) is obtained by starting the derivation from the odd-parity basis of $\sin(n\theta)$ rather than the general angular momentum basis $\exp(in\theta)$.

Nevertheless, we shall use this model for any value of $\tau$ and $k$, as a model which in the resonant and in the generic case (irrational $\tau/(4\pi)$) corresponds to the classical kicked rotator, and in the limit $N\rightarrow \infty$ approaches the infinite dimensional model [Eq.~(\ref{Uoper})], restricted to the symmetry class of the odd eigenfunctions. It is of course just one of the possible discrete approximations to the continuous infinite dimensional model.

The difference of behavior between the generic case and the quantum resonance shows up only at very large times, which grow fast with $(2N+1)$, as explained
in \cite{ManRob2013}.  It turns out that also the eigenfunctions and the spectra of the eigenphases at finite dimension $N$ of the matrices that we consider do not show any significant differences in structural behavior for the rational or irrational $\tau/(4\pi)$, which we have carefully checked. Indeed, although the eigenfunctions and the spectrum of the eigenphases exhibit {\em sensitive dependence on the parameters} $\tau$ and $k$, their statistical properties are stable against the small changes of $\tau$ and $k$. This is an advantage, as instead of using very large single matrices for the statistical analysis, we can take a large ensemble of smaller matrices for values of $\tau$ and $k$ around some central value of $\tau=\tau_0$ and $k=k_0$, which greatly facilitates the numerical calculations  and improves the statistical significance of our empirical results. Therefore our approach is physically meaningful. Similar approach was undertaken by Izrailev (see \cite{Izr1990}
and references therein). In Fig.~1 of paper \cite{ManRob2013} we show the examples of strongly exponentially localized eigenstates by plotting the natural logarithm of the probabilities  $w_n=|u_n|^2$ versus the momentum quantum number $n$, for two different matrix dimensions $N$. By calculating the localization length $\linf$ from the slopes $\sigma=2/\linf$ of these eigenfunctions using Eq.~(\ref{exploc}) we can get the first quantitative empirical localization measure to be discussed and used later on. The new finding of this paper is that $\sigma$ has a distribution, which is close to the Gaussian (but cannot be exactly that, because $\sigma$ is a positive
definite quantity). It does not depend on $N$ and survives the limit $N\rightarrow \infty$. Therefore also $\linf$ has a distribution whose variance does not vanish in the limit $N\rightarrow \infty$.

Following \cite{ManRob2013} and \cite{Izr1990} we introduce another measure of localization.
For each $N$-dimensional eigenvector of the matrix $U_{nm}$ the information entropy is
\be  \label{infoentr}
 \mathscr{H}_N(u_1,...,u_N) = -\sum_{n=1}^{N}w_n \ln w_n,
\ee
where $w_n = |u_n|^2$, and $\sum_n |u_n|^2 = 1$. We denote

\begin{flalign} \label{H_GOE}
 \mathscr{H}_{N}^{GOE}=\psi \left (\frac{1}{2}N+1 \right )-\psi \left (\frac{3}{2} \right )\simeq \ln \left (\frac{1}{2}Na \right )+O(1/N),
\end{flalign}
where $a=\frac{4}{\exp(2-\gamma)}\approx 0.96$, while $\psi$ is the digamma function and $\gamma$ the Euler constant ($\simeq 0.57721...$). 
We thus define the {\it entropy localization length} $l_H$ as 
\be\label{lh:eq}
 l_H=N \exp \left (\mathscr{H}_{N}-\mathscr{H}_{N}^{GOE} \right ).
\ee
Indeed, for entirely extended eigenstates $l_H=N$. Thus, $l_H$ can be calculated for every eigenstate individually. However, all eigenstates,
while being quite different in detail, are exponentially localized,
and thus statistically very similar. Therefore, in order to minimize the fluctuations one uses the {\it mean localization length} $d\equiv \langle l_H \rangle$, which is computed by averaging the entropy over all eigenvectors of the same matrix (or even over an ensemble of similar matrices of the same $N$ but
nearby $k$)
\be\label{d:eq}
 d \equiv \langle l_H \rangle = N \exp \left (\langle \mathscr{H}_{N} \rangle-\mathscr{H}_{N}^{GOE} \right ).
\ee
The {\it localization parameter} $\blo$ is then defined as
\be\label{beta_loc:eq}
    \blo=\frac{d}{N}\equiv \frac{\langle l_H\rangle}{N}.
\ee
The parameter that determines the transition from weak to strong quantum chaos is neither the strength parameter $k$ nor the localization length $\linf$, but the ratio of the localization length $\linf$ to the size $N$ of the system in momentum $p$
\be\label{MLL:Lamda}
  \La=\frac{\linf}{N} = \frac{1}{N}
\left( \frac{\alpha_{\mu} D_{\mu} (K)}{\tau^2}\right)^{\frac{1}{2-\mu}} ,
\ee
where $\linf \approx t_{\rm loc}$, the theoretical localization length
Eq.~(\ref{finallinf}), was derived in \cite{ManRob2013}. $\La$ is the scaling parameter of the system.  The relationship of $\La$ to $\blo$ is discussed in section VII of \cite{ManRob2013}.

\section{The distribution of the localization length and other localization measures}
\label{sec3}

\begin{figure}
\center
\includegraphics[width=7.5cm]{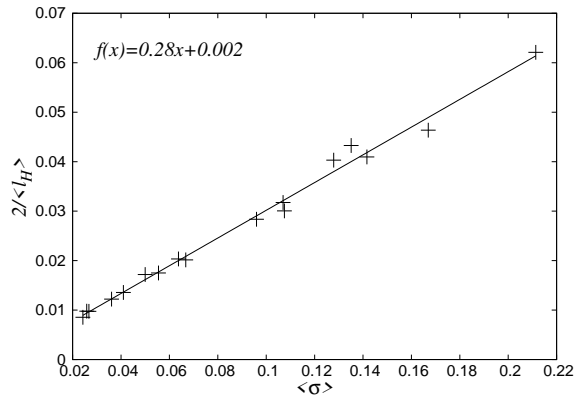}
\caption{We show $\langle\sigma \rangle$ versus $2/\langle l_H \rangle$
for matrices of dimension $N=3000$, for 7 nearby values of $k$, namely $k=k_0 \pm j\delta k$, where $j=0,1,2,3$ and $\delta k=0.00125$, for $k_0=3,4,5,\dots,19$. The two empirical localization measures are clearly well defined, linearly related and thus equivalent.}
\label{figManRob2014-6}
\end{figure}

In this section we present the main results of the paper. First we demonstrate that the localization measures  $2/\sigma$ and $l_H$  are very well defined, linearly related and thus equivalent. In Fig.~\ref{figManRob2014-6} we show this in the  diagram  of the mean $\langle \sigma \rangle$ versus $2/\langle l_H \rangle$, where both averagings are over all eigenfunctions for  matrices of dimension $N=3000$, for 7 nearby values of $k$ around $k_0$, namely $k=k_0 \pm j\delta k$, where $j=0,1,2,3$ and $\delta k=0.00125$, for $k_0=3,4,5,\dots,19$.

In the next Fig.~\ref{figManRob2014-7} we show the relationship of the theoretical $\linf$ in Eq.~(\ref{finallinf}) and the mean value  of the empirical $2/\langle \sigma\rangle$  for  $k_0=3,4,5,...,19$. It is clearly seen in Fig.~\ref{figManRob2014-7}(a) that there are strong fluctuations which we attribute to the fact that $2/\sigma$ has a certain distribution with nonvanishing variance, to be presented and described below, and that the theory of $\linf$ is too simple, as it corresponds only roughly to the value of $2/\langle\sigma\rangle$. On the other hand, in Fig.~\ref{figManRob2014-7} (b) we see again that the two empirical localization measures are exactly linearly related. We should mention that in the cases of larger $k > 19$ the slopes $\sigma$ are so small, and the localization too weak, that we cannot get reliable results, thus in this work we limit ourselves to the interval $3\le k\le19$.

\begin{figure}
\center
\includegraphics[width=7.5cm]{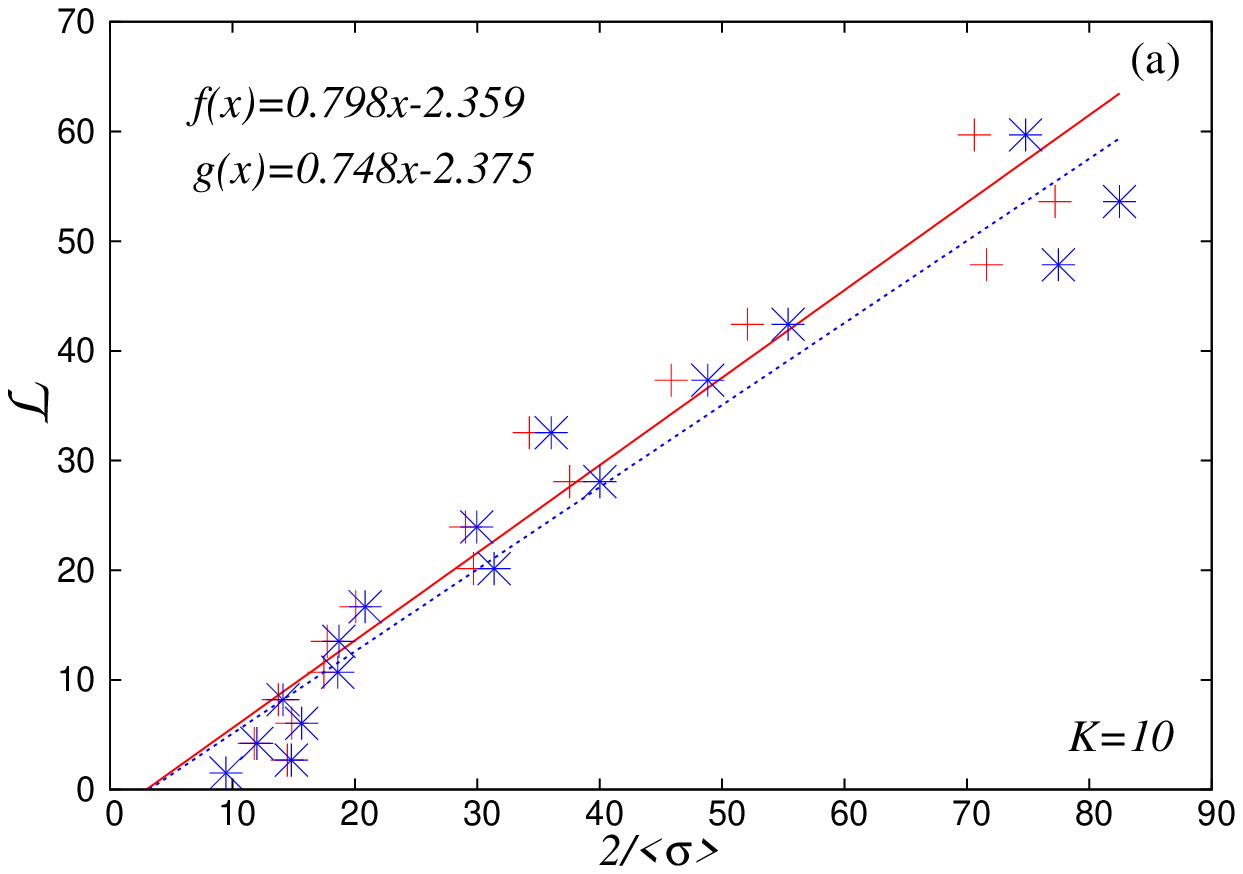}
\includegraphics[width=7.5cm]{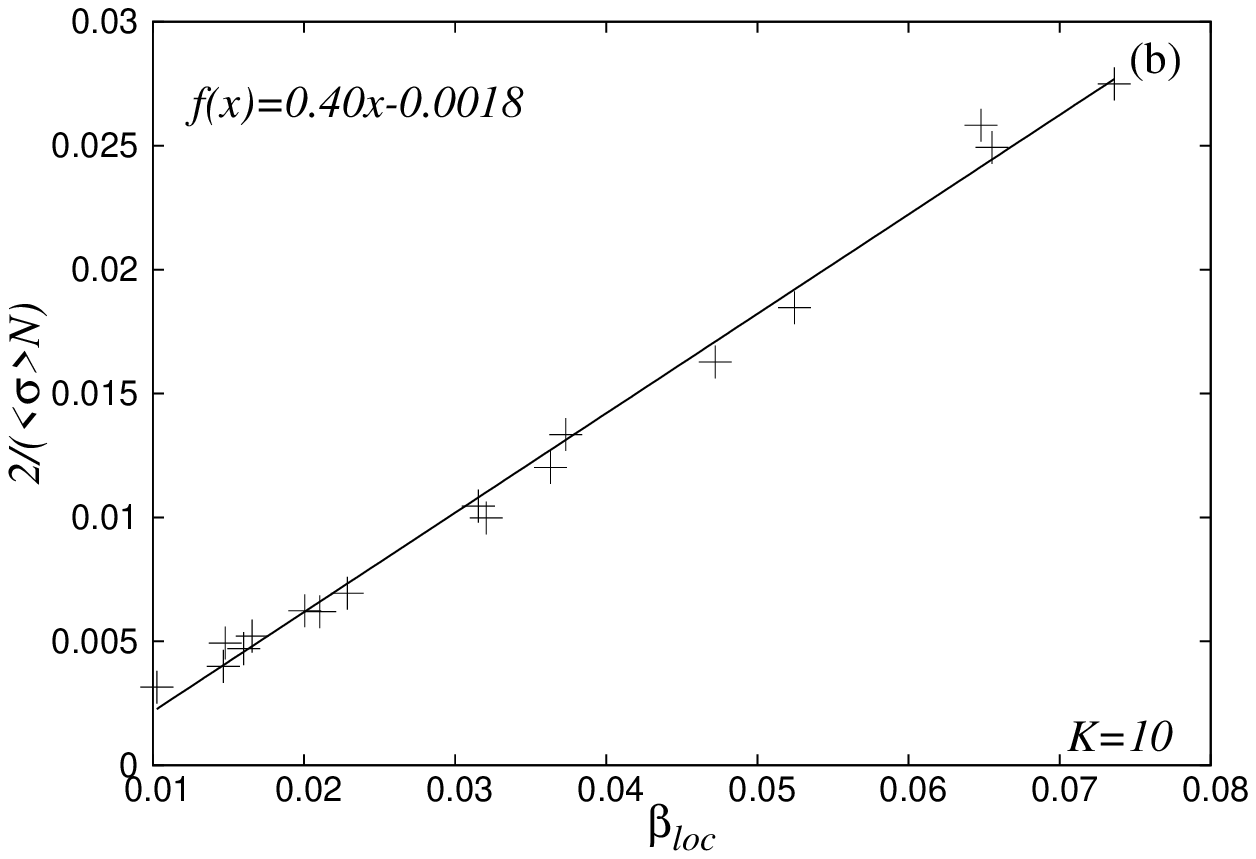}
\caption{[Color online] (a) We show $\linf$ versus $2/\langle \sigma\rangle$
for matrices of dimension $N=1000$ (crosses and solid fit line) and for matrices of dimension $N=3000$ (stars and dashed fit line), for 7 nearby values of $k$, namely $k=k_0 \pm j\delta k$, where $j=0,1,2,3$ and $\delta k=0.00125$, for $k_0=3,4,5,...,19$.  (b) We plot the mean value of $2/(N\langle\sigma\rangle)$ versus $\blo$ for $k_0=3,4,5,\dots,19$ and
7 matrices of dimension $N=3000$ with $k=k_0 \pm j\delta k$,
where $j=0,1,2,3$ and the step size $\delta k=0.00125$.}
\label{figManRob2014-7}
\end{figure}

\begin{figure}
\center
\includegraphics[width=7.5cm]{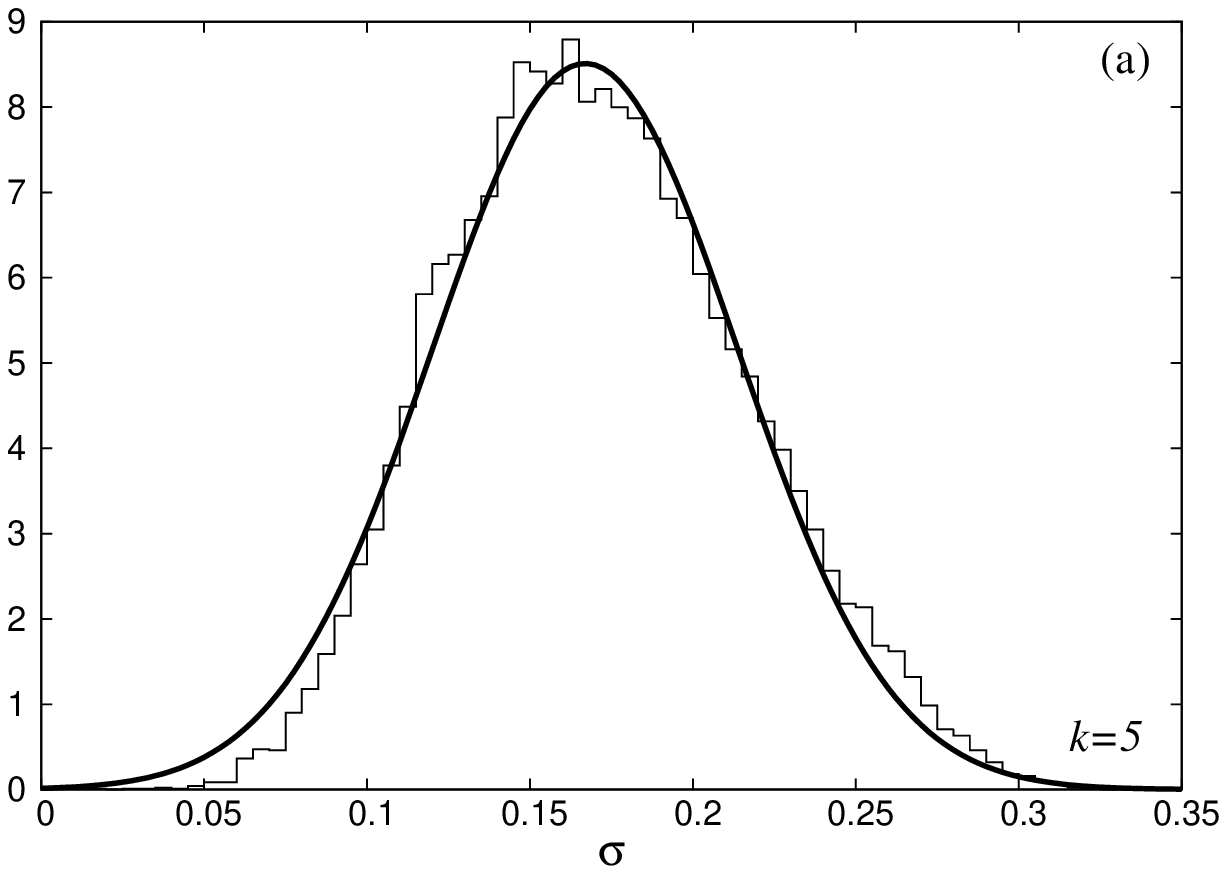}
\includegraphics[width=7.5cm]{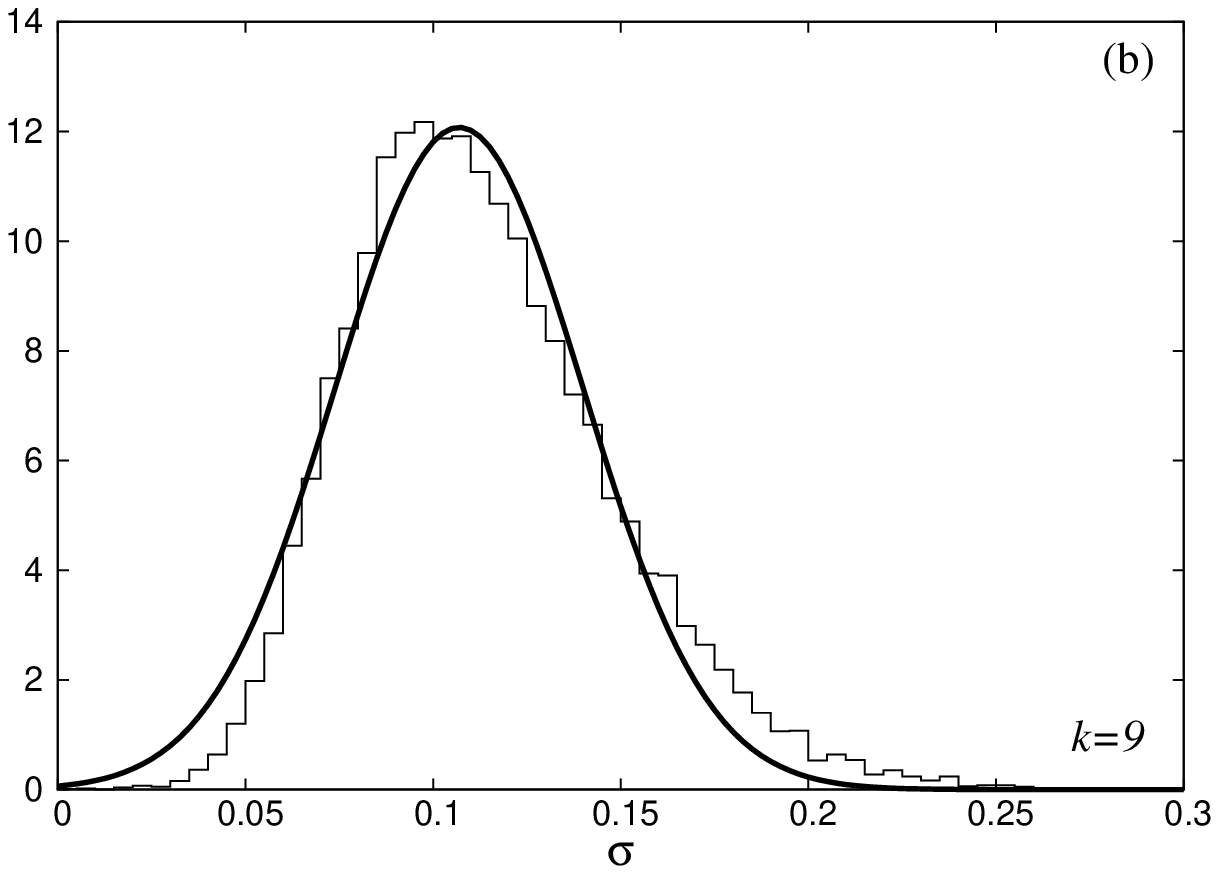}
\includegraphics[width=7.5cm]{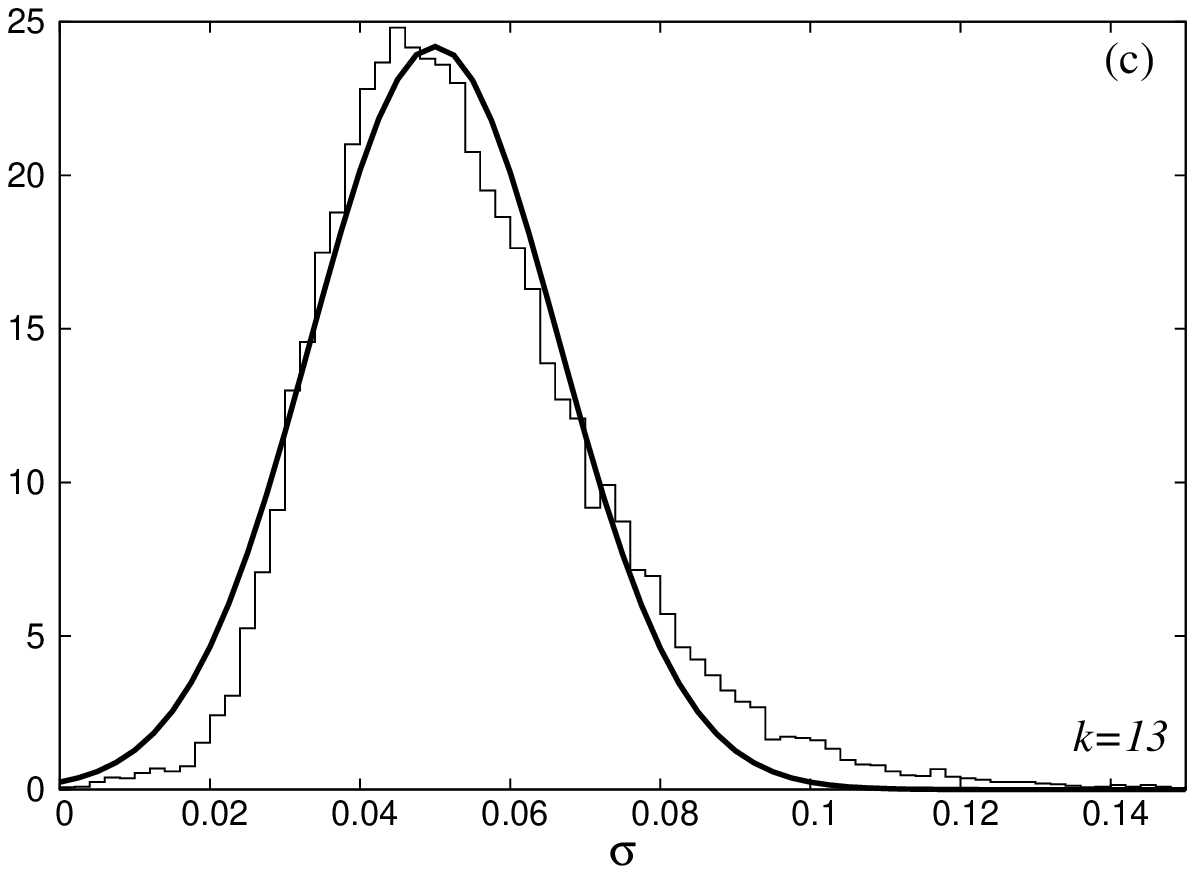}
\includegraphics[width=7.5cm]{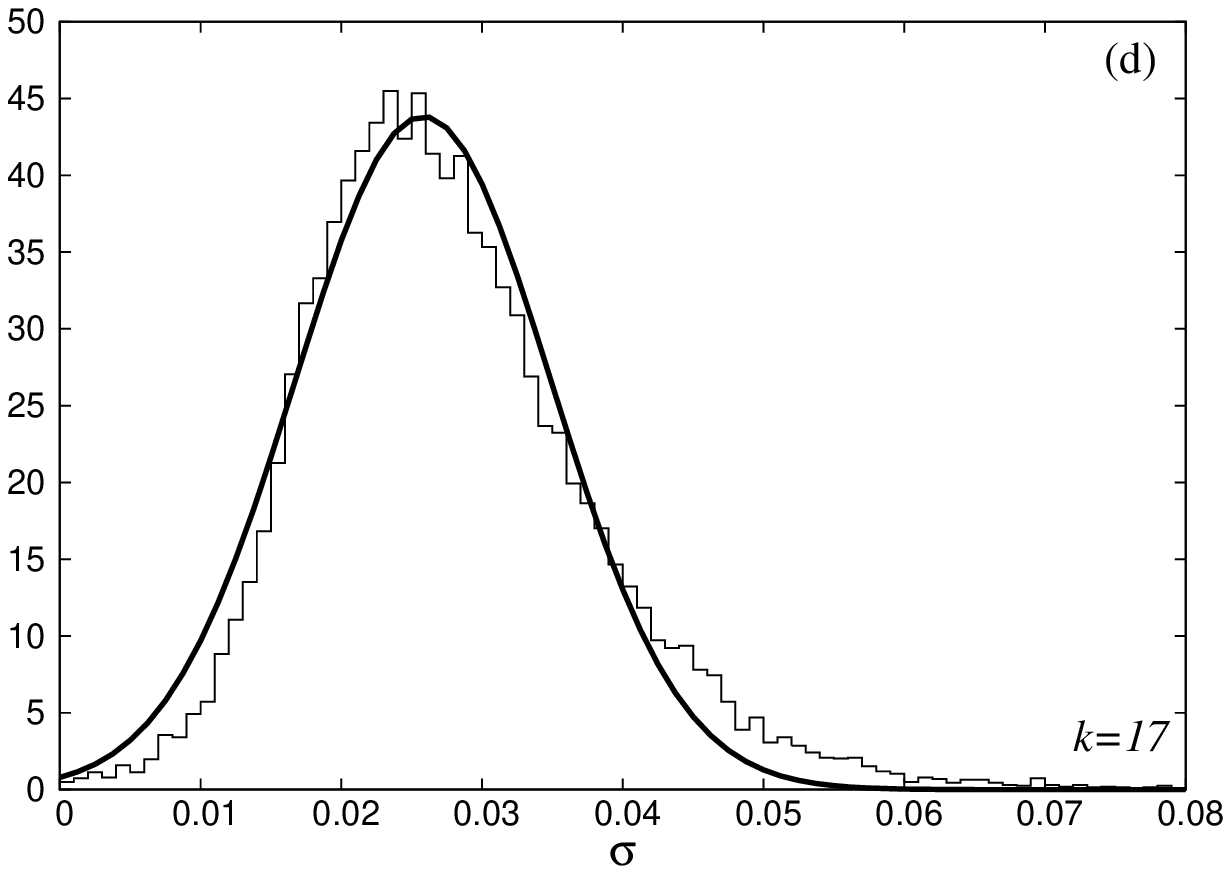}
\caption{We show the histograms of
the slopes $\sigma$ for four systems, matrices of
dimension $N=3000$, for each of them with seven
different values of $k$ close to $k_0=5,9,13,17$,
namely $k=k_0\pm j\delta k$, where $j=0,1,2,3$
and $\delta k=0.00125$:
(a) $k_0=5$, (b) $k_0=9$, (c) $k_0=13$ and (d) $k_0=17$.}
\label{figManRob2014-1}
\end{figure}

Thus we have demonstrated that the empirical localization measures are well defined, while the theoretical prediction for their mean values is not good enough. The reason is that the localization measures of a given fixed system (with fixed $K=10$ and $k$) have a distribution with nonvanishing variance,
which is out of the scope of current semiclassical theories, as they do not predict this distribution and the corresponding variance. This finding as the central result of the present paper is demonstrated in Fig.~\ref{figManRob2014-1}. The distributions are clearly seen to be close to a Gaussian, but cannot be exactly that as $\sigma$ is always a positive definite quantity. Its inverse, the localization length equal to $2/\sigma$, has a distribution whose empirical histograms are much further away from a Gaussian, so that in this sense $\sigma$ is the fundamental quantity. Indeed, as we will see, it corresponds to the  finite time Lyapunov exponent known in the theory of dynamical systems.

\begin{figure*}
\center
\includegraphics[width=7.5cm]{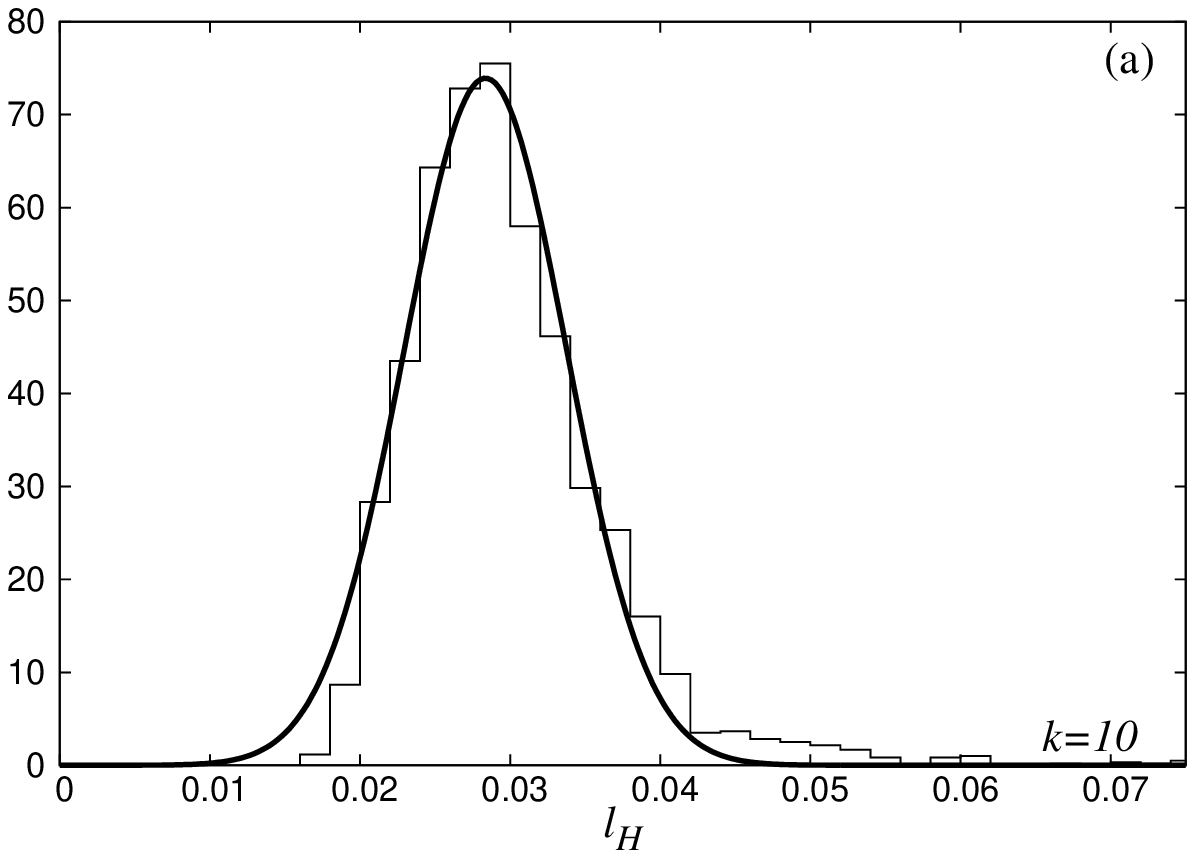}
\includegraphics[width=7.5cm]{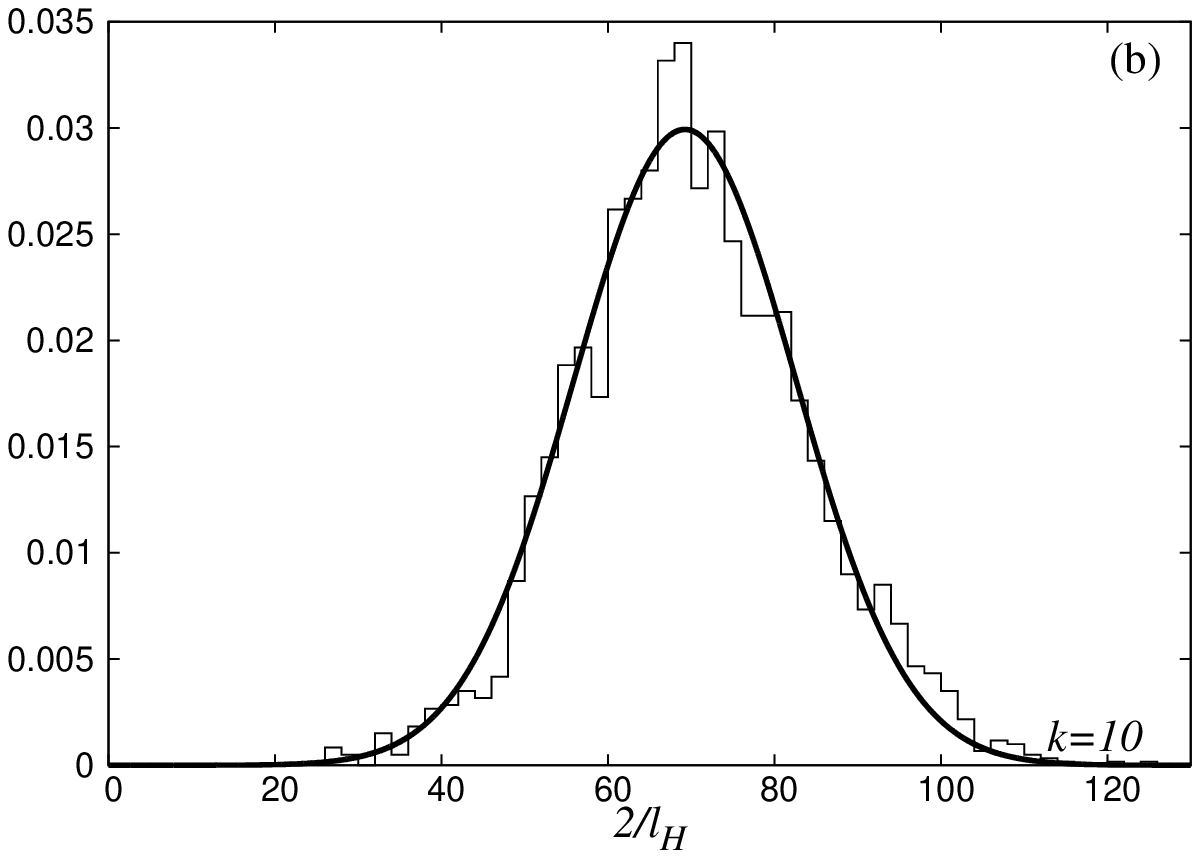}
\caption{We show the histograms of
 $l_H$ in (a) and $2/\l_H$ in (b) for the system
$k=10$ described by the matrices of dimension $N=3000$.
In both cases we show the Gaussian best fit.}
\label{figManRob2014-5}
\end{figure*}

As $l_H$ and  $2/\sigma$ are equivalent localization measures,
the former one is expected also to have a distribution, which
we demonstrate in the histograms of Fig.~\ref{figManRob2014-5}.

\begin{figure*}
\center
\includegraphics[width=7.5cm]{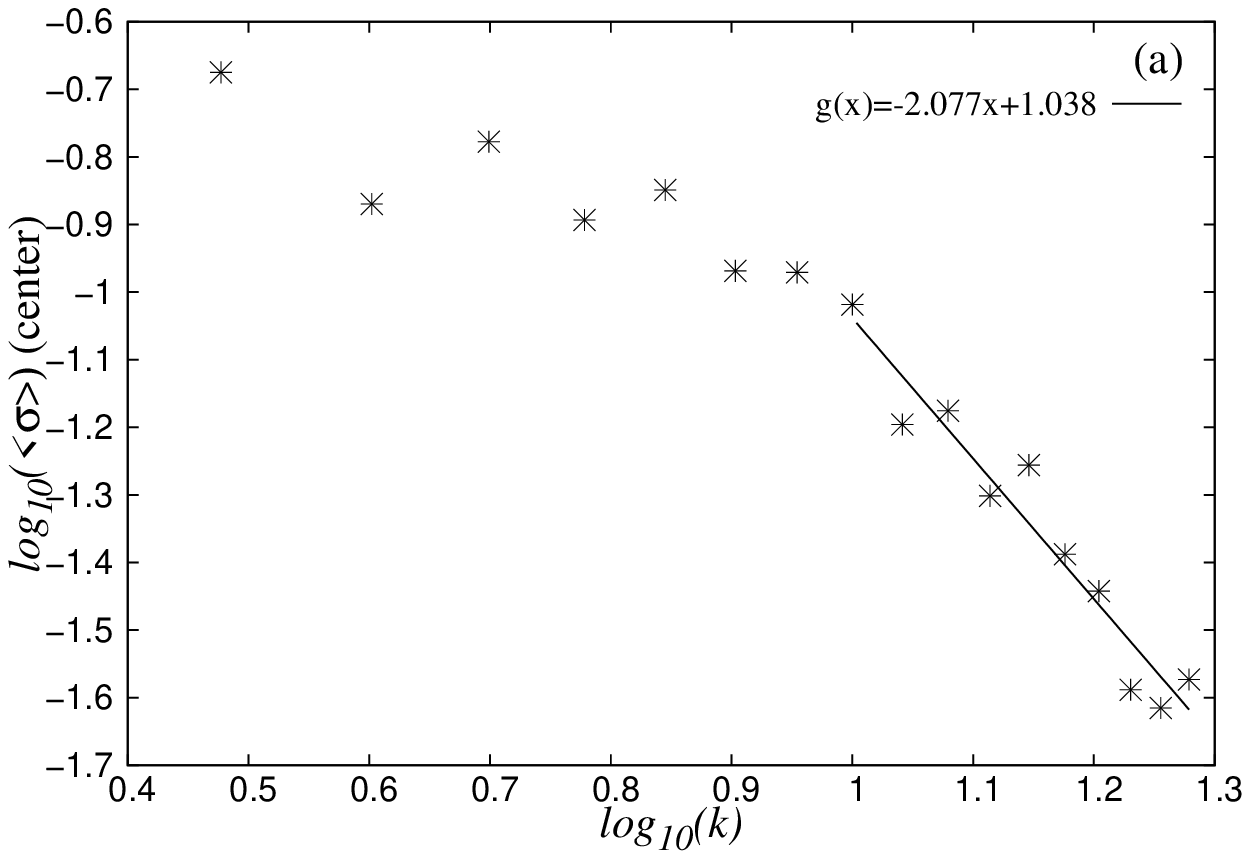}
\includegraphics[width=7.5cm]{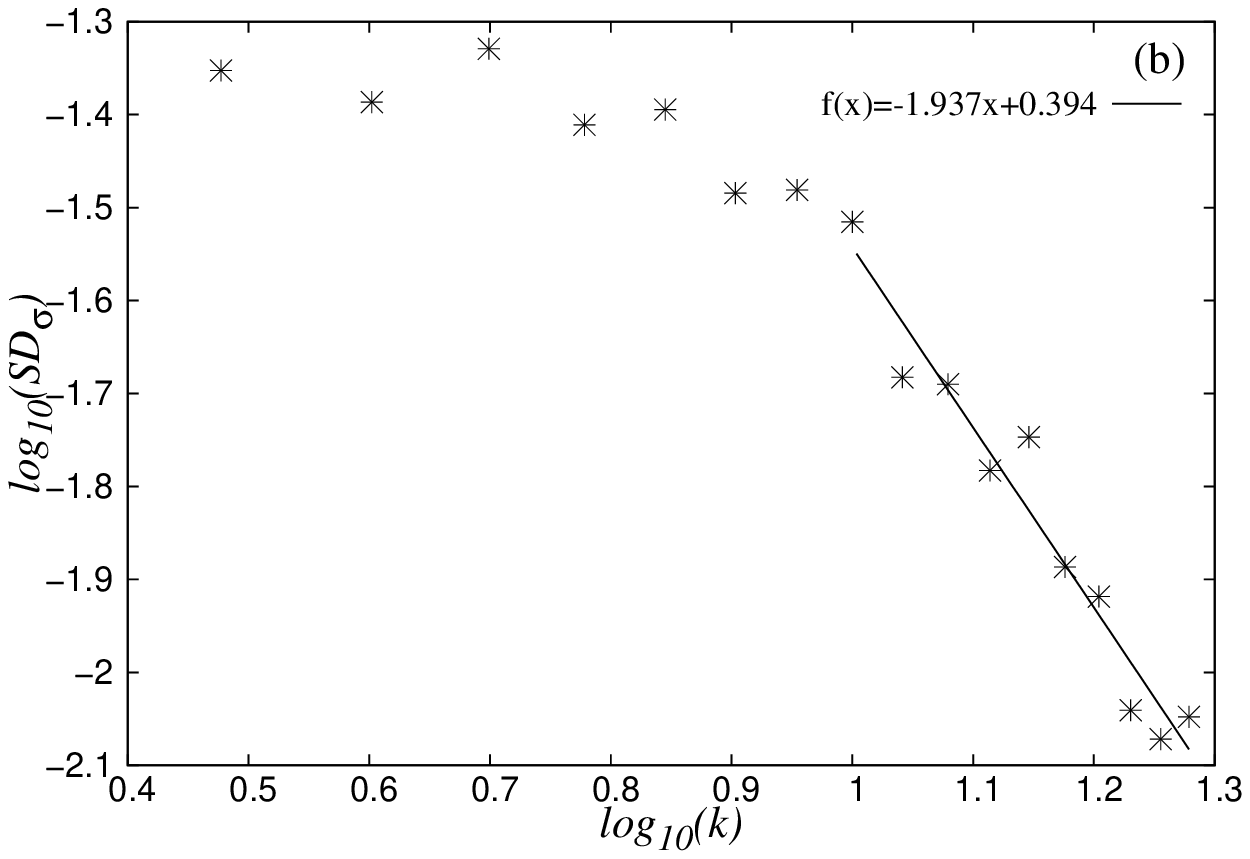}
\caption{We show log-log plots in (a) the mean slope
$\langle \sigma \rangle$ as a function of $k$, and in (b) the standard deviation of $\sigma$ as a function of $k$. The fitting by a straight line is only on the semiclassical interval $10\le k\le 19$. In the former case the behavior is roughly as $1/k^2$, in agreement with the theoretical estimate $1/k^2$ of Eq.~(\ref{finallinf}), and in the latter case also like $1/k^2$, surely not as the theoretical estimate $1/k$ based on the Lyapunov
exponents method in the reference \cite{Kottos1996} [Eq.~(9)].}
\label{ManRob2014newfig5}
\end{figure*}

We have also analyzed how the localization measures vary in the semiclassical limit of the increasing value of the quantum parameter $k$, at fixed classical parameter $K=10$. Indeed, the theoretical estimate of $\linf$ in Eq.~(\ref{finallinf}), at fixed $K$, and remembering $k=K/\tau$, shows that approximately the mean value of the localization length should increase quadratically with $k$, or equivalently, the slope $\sigma$ should decrease inversely quadratically with $k$. This prediction is observed, and is demonstrated in the Table~\ref{tabManRob2014-1}, and also in Fig.~\ref{ManRob2014newfig5}.
%
%
%
It is also in agreement with the prediction based on the tight-binding approximations in reference \cite{Kottos1996} [Eq.~(6)]. We give, in Table~ \ref{tabManRob2014-1}, the mean slope $\sigma$ and the standard deviation of $\sigma$, as well as the mean value of the related quantity $2/l_H$ and its standard deviation for various $k=k_0=3,4,5,\dots,19$, for each of them taking
seven nearby values of $k$, namely $k=k_0\pm j\delta k$, where $j=0,1,2,3$ and $\delta k =0.00125$, for matrices of dimension $N=3000$. Each histogram for all $k_0$ was fitted with the Gaussian distribution and then the mean values and the standard deviations were extracted. All four quantities decrease to zero with increasing $k$, meaning that in the semiclassical limit the localization
lengths monotonically increase to infinity, so that in this limit we have asymptotically extended states (no localization), and their standard deviation also goes to zero as $1/k^2$, which is different from the tight-binding approximations in reference \cite{Kottos1996} [Eq.~(9)].

\begin{table}\caption{\label{tabManRob2014-1}}
\textbf{The mean value and the standard deviation of the slopes
$\sigma$
and $2/l_H$ as a function of $k=k_0=3,4,5,\dots,19$. 
For each $k=k_0$ we used $N=7\times 3000$ slopes $\sigma$ (see text). 
All quantities decay to zero in the semiclassical limit.}
\begin{ruledtabular}
\begin{tabular}{cccccc}
\multicolumn{5}{c}{$K=10$ \ -- \ $N=7\times3000$~(slopes) -- \ $N=3000~(2/l_H)$} \\
$k$ & $<\sigma>$ & $SD_{\sigma}$ & $<2/l_H>$ & $SD_{2/l_H}$\\
\colrule
3  & 0.06209 & 0.01324 & 0.062098 & 0.01324 \\
4  & 0.04327 & 0.01073 & 0.043272 & 0.01073 \\
5  & 0.04636 & 0.00758 & 0.046363 & 0.00758 \\
6  & 0.04030 & 0.00974 & 0.040303 & 0.00974 \\
7  & 0.04095 & 0.00838 & 0.040954 & 0.00838 \\
8  & 0.03004 & 0.00756 & 0.030047 & 0.00756 \\
9  & 0.03174 & 0.00600 & 0.031743 & 0.00600 \\
10 & 0.02835 & 0.00539 & 0.028355 & 0.00539 \\
11 & 0.02034 & 0.00353 & 0.020341 & 0.00353 \\
12 & 0.02014 & 0.00321 & 0.020143 & 0.00321 \\
13 & 0.01719 &  0.0029 & 0.017193 & 0.00294 \\
14 & 0.01750 & 0.00289 & 0.017509 & 0.00289 \\
15 & 0.01356 & 0.00230 & 0.013569 & 0.00230 \\
16 & 0.01221 & 0.00194 & 0.012213 & 0.00194 \\
17 & 0.00978 & 0.00148 & 0.009787 & 0.00148 \\
18 & 0.00855 & 0.00128 & 0.008550 & 0.00128 \\
19 & 0.00975 & 0.00141 & 0.009754 & 0.00141 \\
\end{tabular}
\end{ruledtabular}
\end{table}

Next we want to study how does the distribution of the localization measure
$\sigma$ behave as a function of the dimension $N$ of the Izrailev model
Eqs.~(\ref{u_repres}-\ref{Bnmoper}). Since in the limit $N\rightarrow \infty$ the model converges to the infinitely dimensional quantum kicked rotator, we would
at first sight expect that following the Shepelyansky picture \cite{She1986}
$\sigma$ should converge to its asymptotic value, which is sharply defined
in the sense that the variance of the distribution of $\sigma$ goes to zero
inversely with $N$. Namely, at fixed $K$ and $k$ Shepelyansky reduces the problem of calculating the localization length to the problem of the finite time Lyapunov exponents of the {\bf approximate} underlying finite dimensional Hamilton system with dimension $2k$. The localization length is then found to be equal to the inverse value of the smallest positive Lyapunov exponent.  In our case, the dimension of the matrices $N$ of the Izrailev model plays the role of time.  As it is known, and analyzed in detail in the next sections, the finite time Lyapunov exponents have a distribution, which is almost Gaussian, and its variance decays to zero inversely with time. Thus on the basis of
this we would expect that the variance of $\sigma$ decays inversely with $N$.

However, this is not what we observe. In the Table~\ref{tabManRob2014-2}
we clearly see that at constant $K=10$ and $k=10$ the mean value of $\sigma$
is constant and obviously equal to its asymptotic value of $N=\infty$,
while the variance of $\sigma$ does not decrease with $N$, as $1/N$,
but is constant instead, independent of $N$. This is in disagreement with
the banded-matrix models of the tight-binding approximations and thus
disagrees with the Eq.~(9) of reference \cite{Kottos1996}, and also disagrees with the Shepelyansky picture. The reason  is that the associated Shepelyansky's Hamilton system is only approximate construction, because with increasing $N$ the matrix elements of the Floquet propagator (matrix) outside the diagonal band of width $2k$ become important, and thus the dimension of
the Hamilton system cannot be considered finite, constant and equal to $2k$,
but increases with $N$. As a consequence we have the constant value of the variance of $\sigma$, and thus constant variance of the localization length ${\cal L}=2/\sigma$, and therefore the localization length has a distribution with nonvanishing variance even in the limit $N=\infty$. This is precisely the reason why the semiclassical prediction of the localization length in
Eq.~(\ref{finallinf}) fails in detail and we find strong fluctuations in the plot of $\linf$ against the $2/\sigma$ of Fig.~\ref{figManRob2014-7}. The proper theory of the localization length must predict its distribution rather than just its approximate mean value.

\begin{table}\caption{\label{tabManRob2014-2}}
\textbf{The mean value and the variance of the slope $\sigma$ as
a function of the matrix dimension $N$ for a fixed system
with $K=10$ and $k=10$. Both are obviously constant.}
\begin{ruledtabular}
\begin{tabular}{cccccc}
\multicolumn{3}{c}{$K=10$ \ -- $k=10$ } \\
$N$ & $\langle \sigma \rangle$ & $var_{\sigma}$ \\
\colrule
500   & 0.102624 & 0.00113224 \\
1000  & 0.101170 & 0.00112558 \\
2000  & 0.100066 & 0.00115575 \\
3000  & 0.102217 & 0.00110438 \\
\end{tabular}
\end{ruledtabular}
\end{table}

\section{Numerical study of finite time Lyapunov exponents
for the classical standard map} \label{sec4}

Finite time Lyapunov exponents of chaotic systems is a subject of not very much intense research. Taking an ensemble of uniformly distributed initial conditions of a uniformly chaotic (ergodic) system (with no islands of stability) we of course expect that for any finite time the Lyapunov exponents will have a certain distribution. With increasing time the  mean value of each of them is expected to converge to the asymptotic Lyapunov exponent, and since the asymptotic Lyapunov exponent must be the same for all initial conditions,
the distribution must converge to the Dirac delta distribution. Some early results on this topic go back to the 1980s, in the works of Fujisaka \cite{Fuj1983}, reviewed and summarized by Ott \cite{Ott1993}. Some details are not so important, as it turns out that the distribution becomes Gaussian very fast with increasing time, which we want to demonstrate in this section.

\begin{figure}
\center
\includegraphics[width=7.5cm]{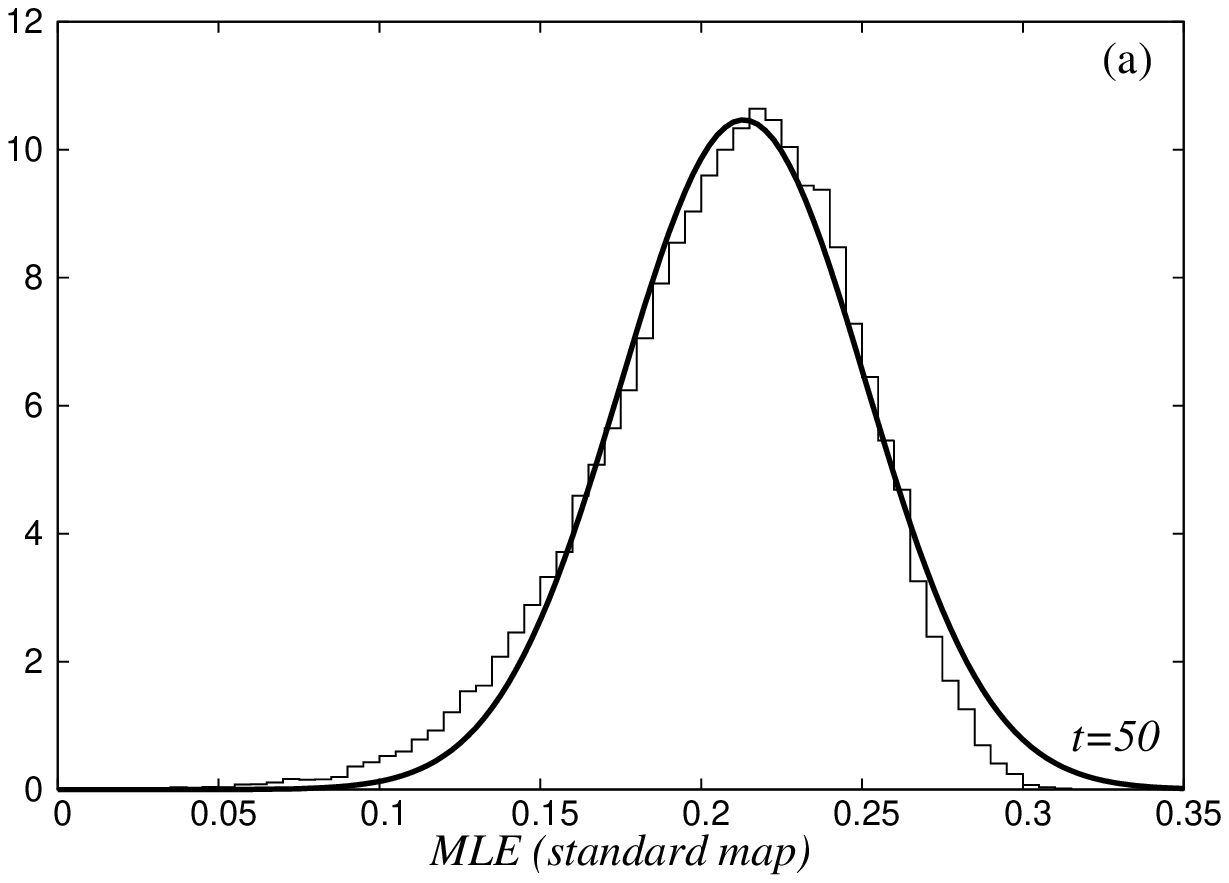}
\includegraphics[width=7.5cm]{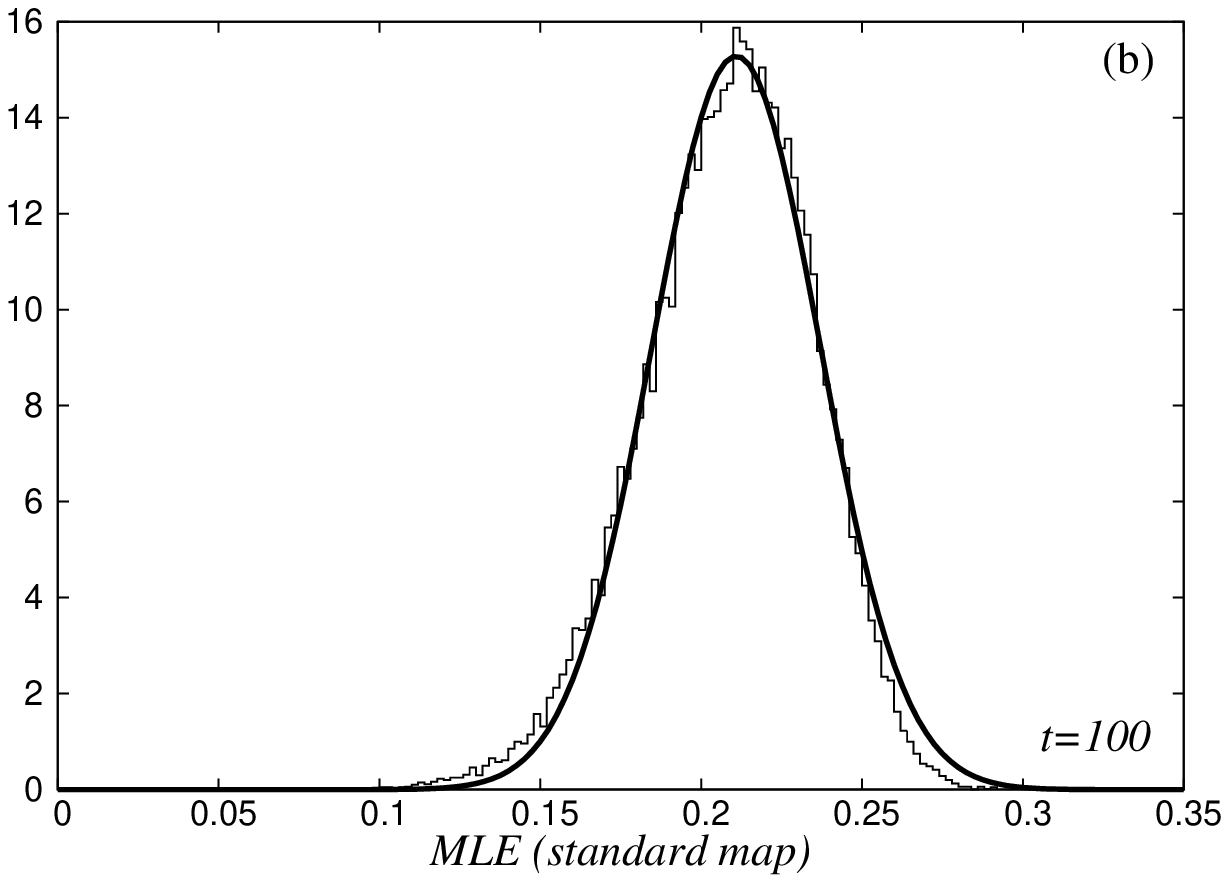}
\includegraphics[width=7.5cm]{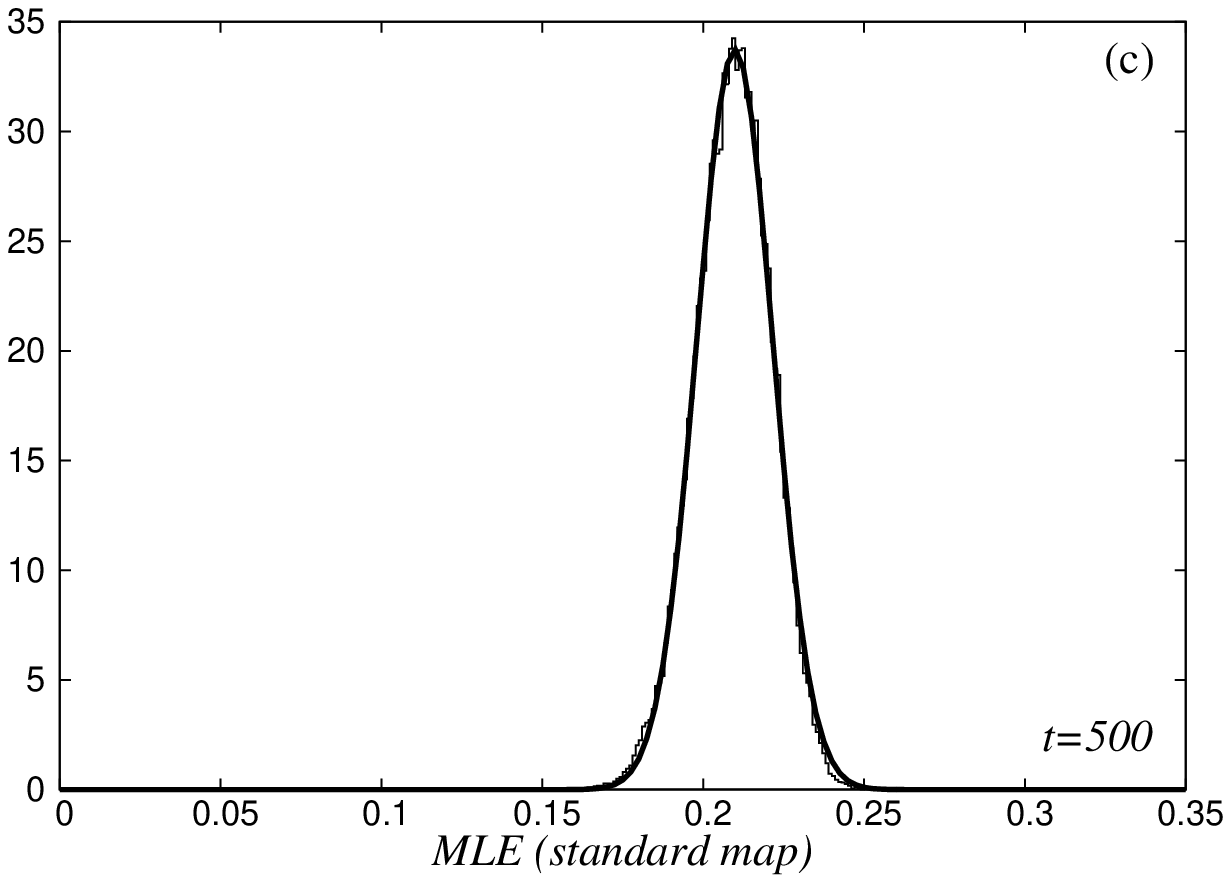}
\includegraphics[width=7.5cm]{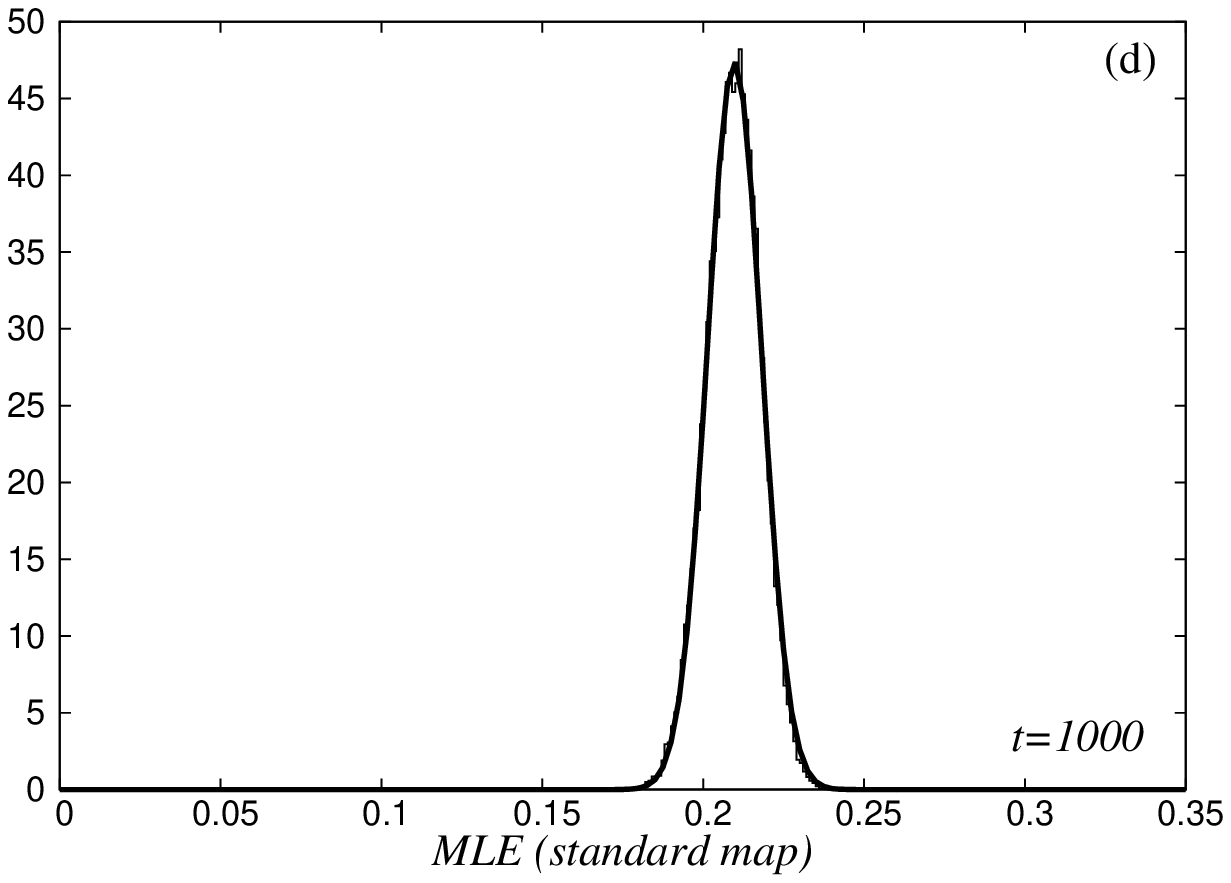}
\caption{We show the histograms of the positive finite time Lyapunov exponents for the standard map [Eq.~(\ref{SM2})] with $K=10$ and times (number of iterations) $t=50,100,500,1000$ in (a), (b), (c) and (d) respectively. The initial conditions are on the grid $200 \times 200$ on the square $[0,2\pi)\times [0,2\pi)$. In all cases we show the Gaussian best fit.}
\label{figManRob2014-2}
\end{figure}

In Fig.~\ref{figManRob2014-2} we show the histograms of the positive finite time Lyapunov exponent for the standard map [Eq.~(\ref{SM2})] with $K=10$, for the finite times (=number of iterations) $t=50,100,500,1000$ in (a), (b), (c) and (d), respectively. The initial conditions, $200\times 200$ on a grid, have been taken uniformly distributed over the square $2\pi \times 2\pi$. Already at $t=50$ the distribution is quite close to a Gaussian, and this trend increases very fast. At longer times like $t=2000, 3000, 4000, 5000$ it becomes a perfect Gaussian distribution (not shown). The variance decreases as $1/t$, as it is demonstrated and analyzed in Sec.~\ref{sec6}.

\section{Numerical study of finite time Lyapunov exponents
for the product of random symplectic 2D matrices} \label{sec5}

\begin{figure}
\center
\includegraphics[width=7.5cm]{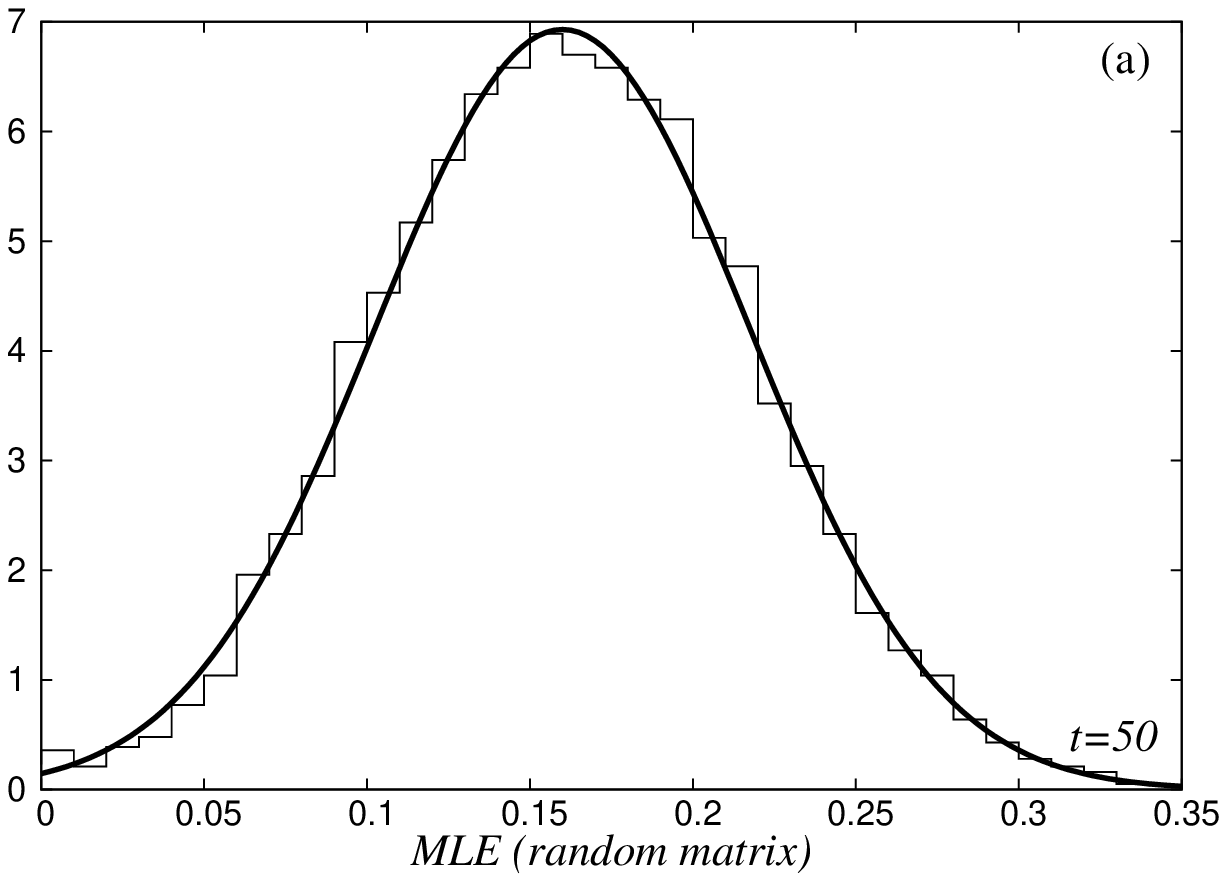}
\includegraphics[width=7.5cm]{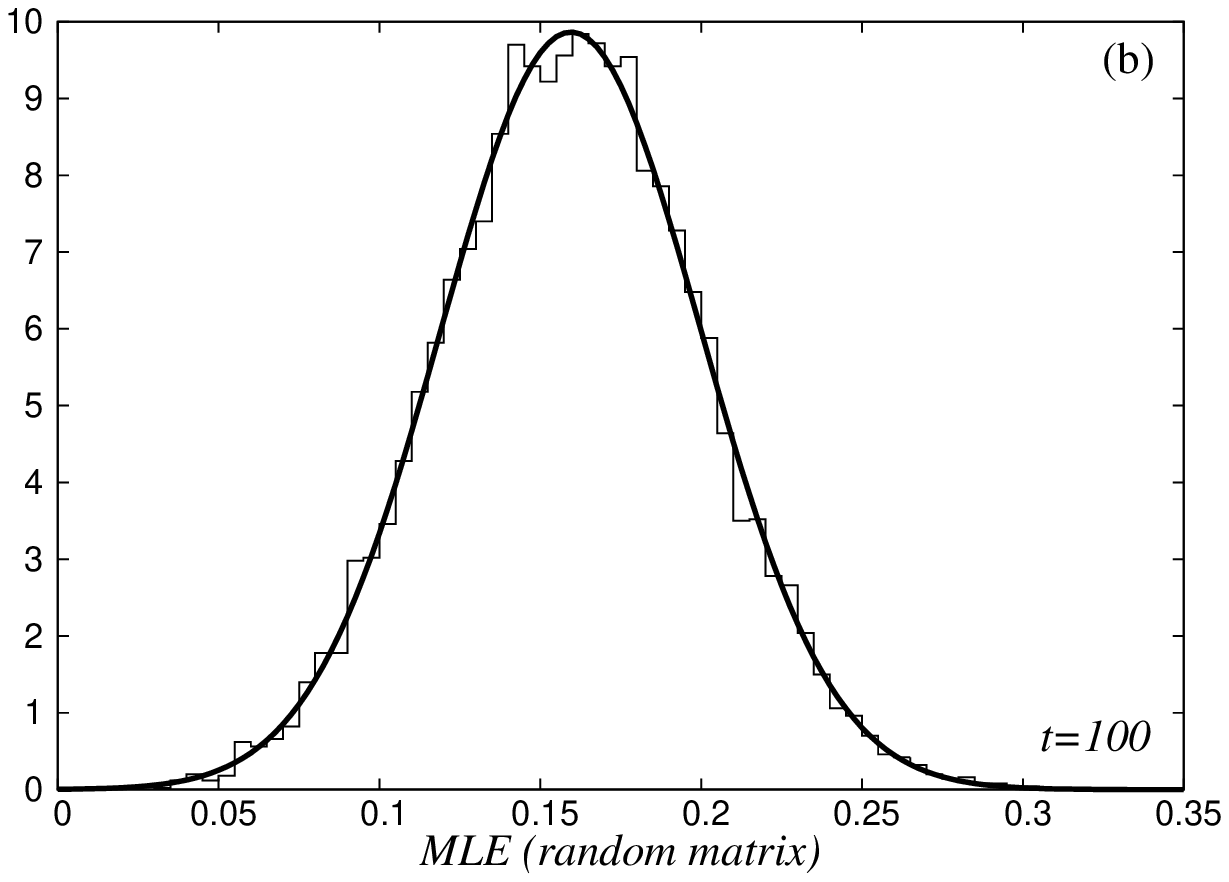}
\includegraphics[width=7.5cm]{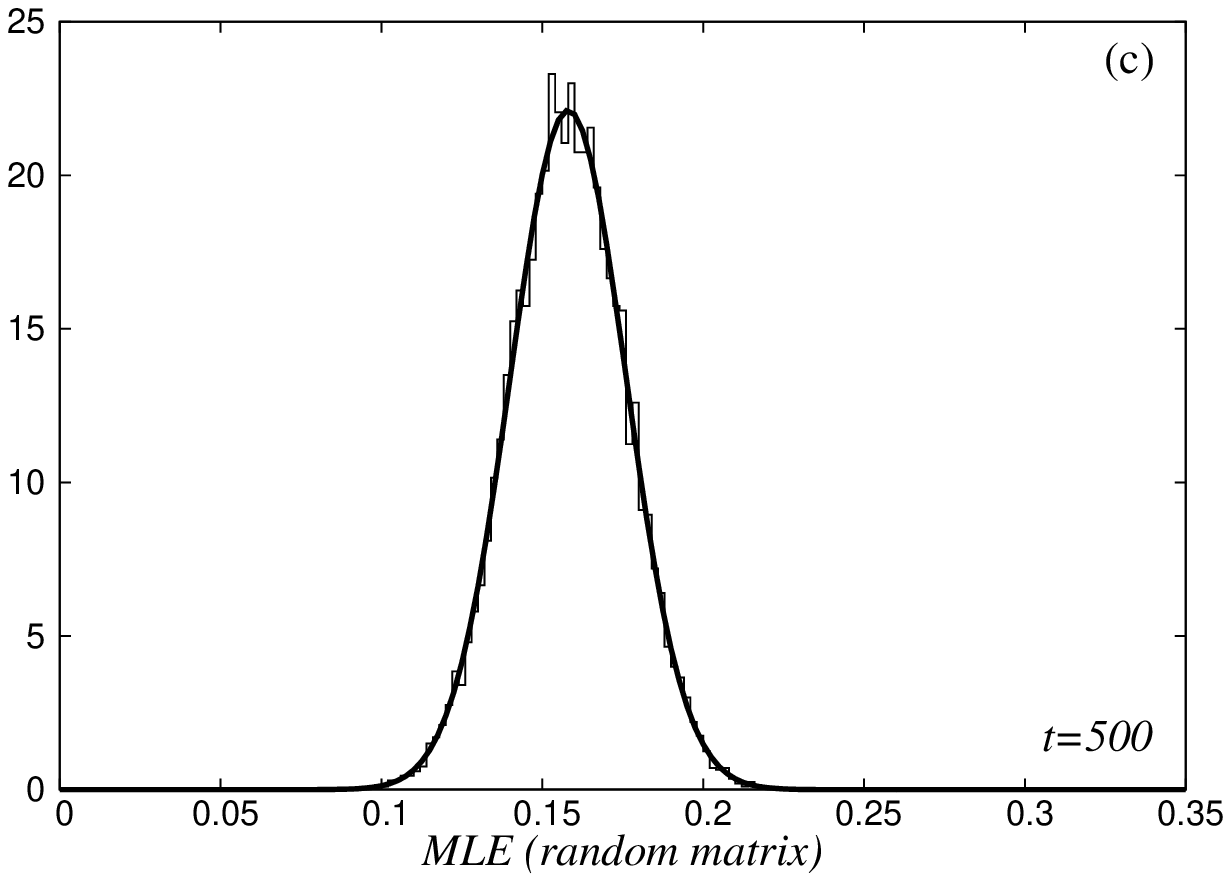}
\includegraphics[width=7.5cm]{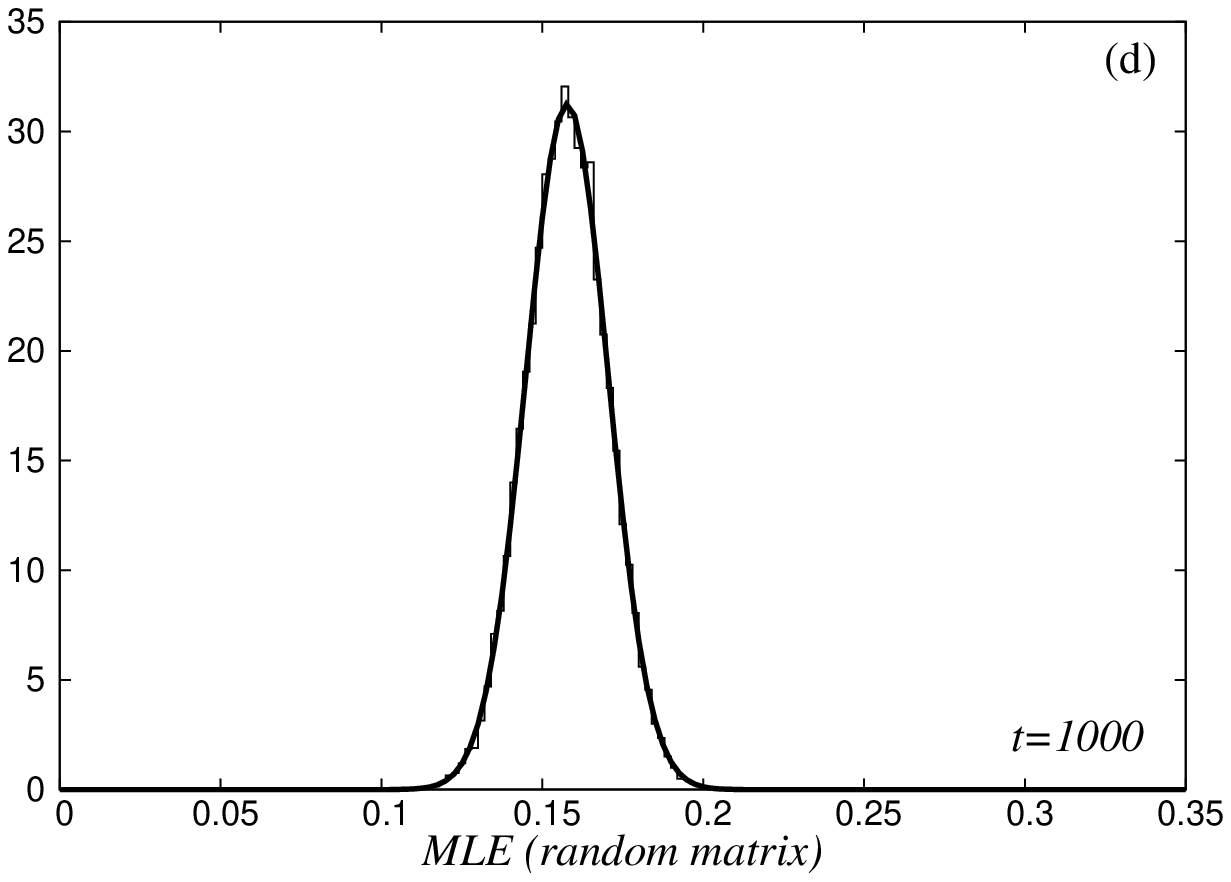}
\caption{We show the histograms of the positive finite time Lyapunov exponents for the product of random matrices [Eq.~(\ref{transfermat})] with $W=E-E_n^0$ uniformly distributed in a box $W\in[-2,+2]$, for $n=50,100,500,1000$ in (a), (b), (c) and (d) respectively. In all cases we show the Gaussian best fit.}
\label{figManRob2014-3}
\end{figure}

As it is well known the problem of quantum or dynamical localization is related to the Anderson localization model, within the framework of the tight-binding approximation, with hopping transitions between the nearest neighbors only. This goes back to the pioneering work of Fishman, Grempel and Prange \cite{FGP1982}, as discussed in \cite{Haake,Stoe}, and also reviewed in \cite{PGF1984}. Assuming the nonvanishing nearest neighbor interaction only and the site disorder, the governing Schr\"odinger equation is \cite{Stoe}
\begin{flalign} \label{Anderson1}
a_{n+1} + E_n^0a_n +a_{n-1} = E a_n
\end{flalign}
where $E$ is the eigenenergy of the eigenfunction, while $E_n^0$ is the fluctuating on-site potential, varying from site to site, with a certain probability distribution. Therefore we have the equation
\be \label{Anderson2}
\left(
\ba {c}
a_{n+1} \\ a_n
\ea
\right)
= T_n \left(
\ba {c}
a_{n} \\ a_{n-1}
\ea
\right)
\ee
where the $2\times2$ transfer matrix $T_n$ is given by
\be \label{transfermat}
T_n =
\left(
\ba {cc}
E-E_n^0 & -1 \\
1 & 0
\ea
\right)
\ee
The determinant is equal to one, and $W=E-E_n^0$ is drawn from a distribution, defined by a given model. Therefore the asymptotic properties of the eigenfunction coefficients $a_n$ as a function of $n$ are determined by the behavior of the product of the random transfer matrices, $T=T_nT_{n-1}\dots T_2T_1$. Everything is determined by the trace $B=Tr T$. If $|B|>2$
the eigenvalues of $T$ are real reciprocals, $\lambda>1$ and $1/\lambda<1$.
Typically $\lambda$ grows exponentially with $n$, and $M_n=n^{-1} \ln \lambda > 0$ fluctuates with $n$, has certain distribution for each finite $n$, and
the limit $M= \lim_{n\rightarrow \infty} M_n$ exists. The latter is known as Furstenburg theorem \cite{CPV1993}. Thus, for generic initial condition
$(a_0,a_1)$ the $a_n$ will grow exponentially with $n$ and only for a special initial condition, they will decrease exponentially with the rate $M_n$ as $n\rightarrow +\infty$, but still will increase exponentially in the backward direction $n\rightarrow -\infty$.  There are  then exactly the eigenenergies $E$ for which the $a_n$ decrease exponentially in both directions $n\rightarrow \pm \infty$. In such case then $M_n$, {\em the finite time Lyapunov exponent}, is precisely the inverse value of the localization length in the $n$-space.
Thus, for the finite system $n< \infty$, we shall have a certain distribution of the Lyapunov exponents $M_n$. Indeed, this is observed in our numerical experiments shown in Fig.~\ref{figManRob2014-3}, for the box distribution of $W$, namely within the interval $W\in [-2,+2]$, for four values $n=50,100,500,1000$, and for each of them for 10~000 realizations, drawn from the distribution of $W$, and we see that the Gaussian approximation is very good, and becomes perfect for longer values of $n$, such as
$n=2000,3000,4000,5000$ (not shown).

\begin{figure}
\center
\includegraphics[width=7.5cm]{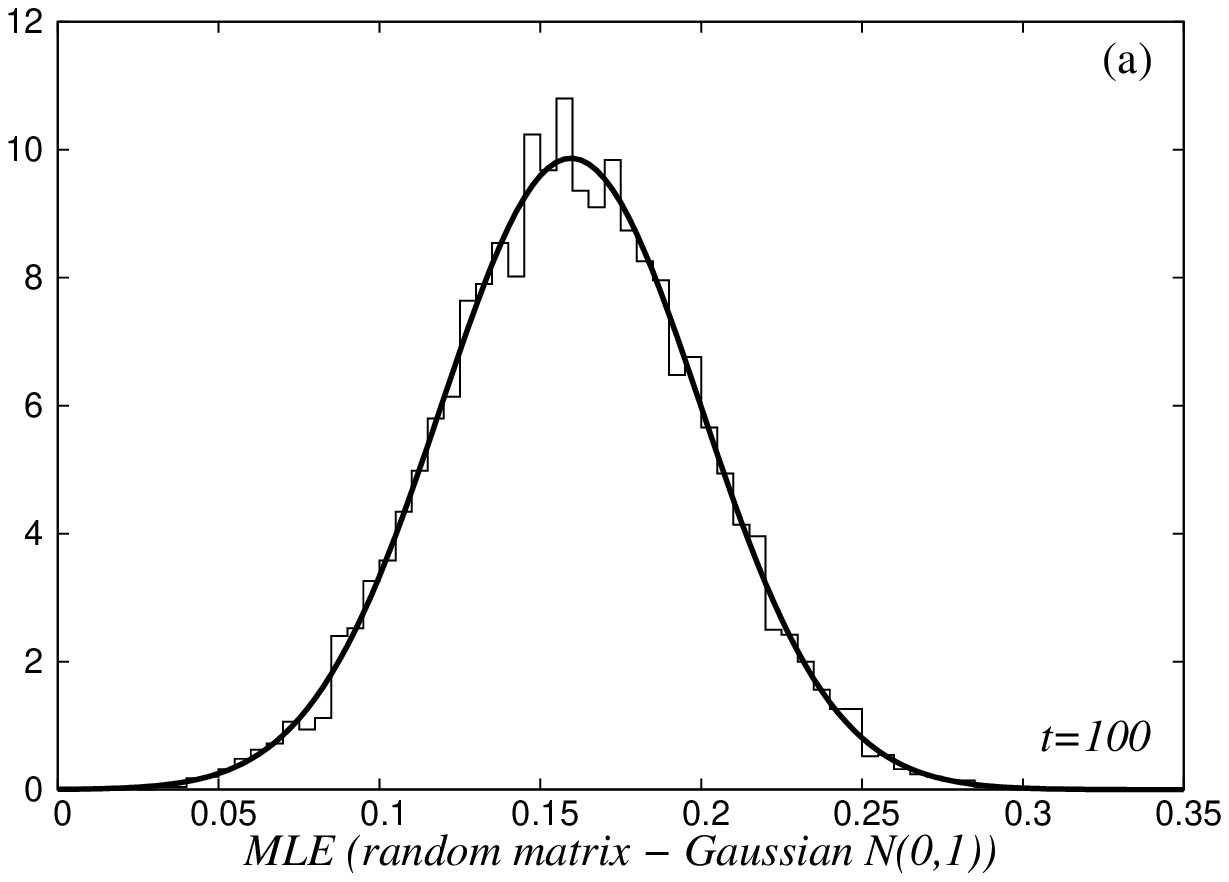}
\includegraphics[width=7.5cm]{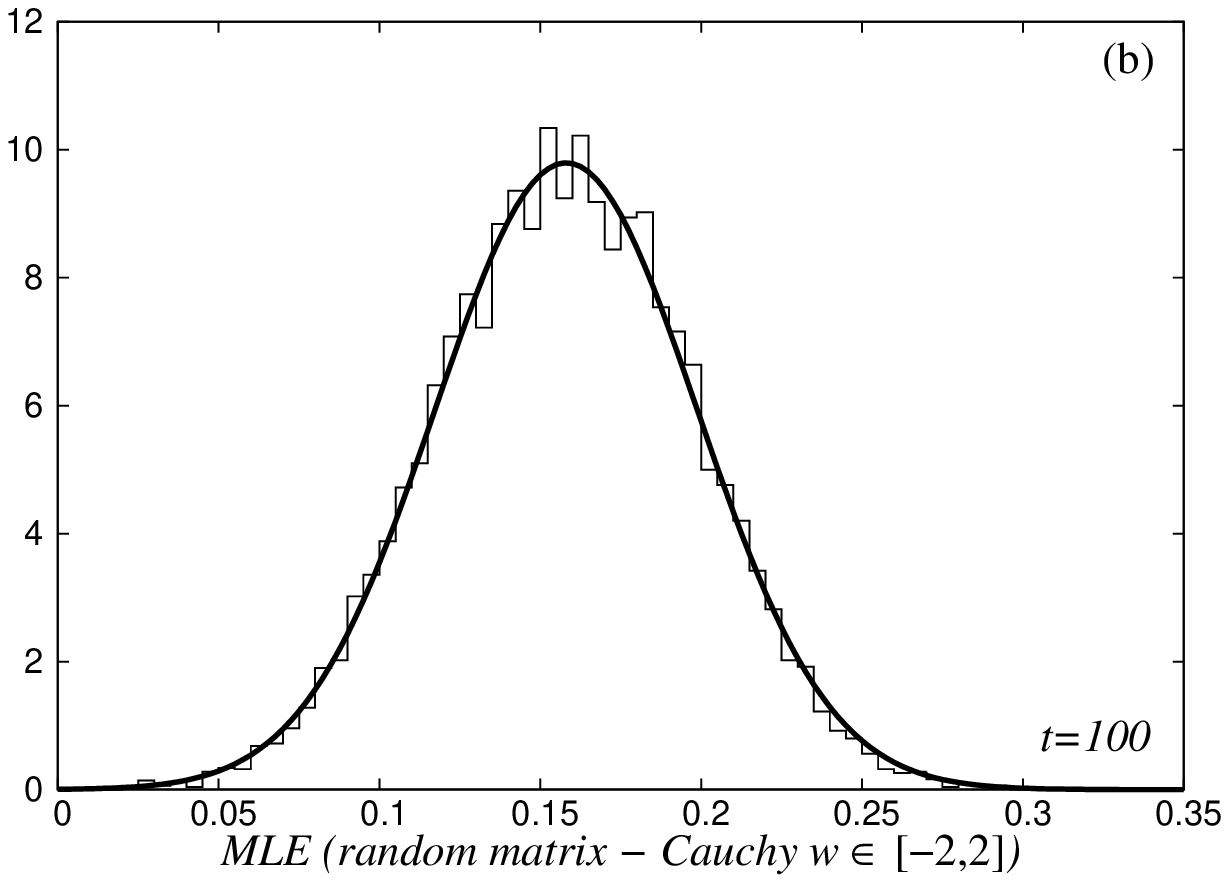}
\includegraphics[width=7.5cm]{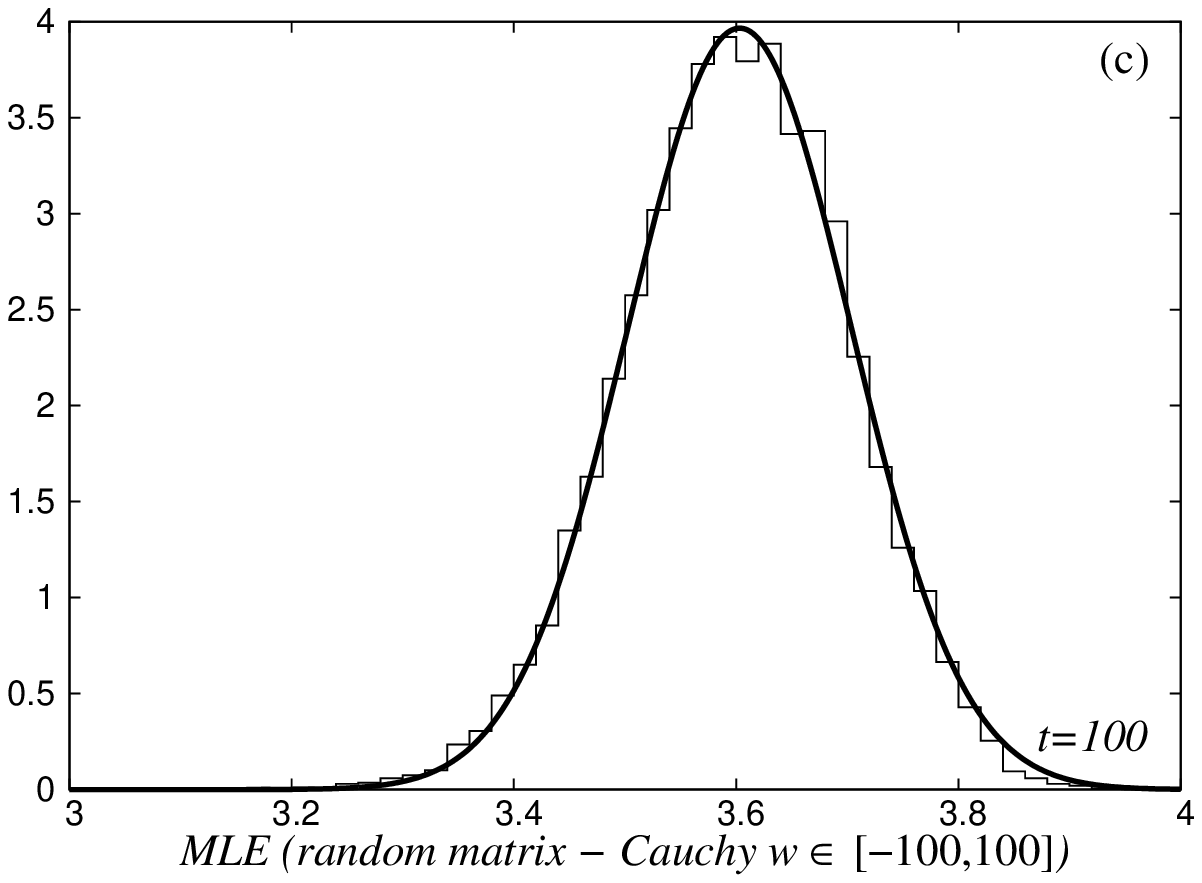}
\caption{The histograms of the positive finite time Lyapunov exponents for the
product of random matrices [Eq.~(\ref{transfermat})] at $n=100$  with (a) $W=E-E_n^0$ Gaussian distributed with zero mean and unit variance, (b) Cauchy-Lorentz distribution [Eq.~(\ref{Cauchy-Lorentz})] with $W$ in the cut-off interval $[-2,+2]$, and (c) the same as (b) but $W\in[-100,+100]$. In all cases we show the Gaussian best fit which is excellent.}
\label{figManRob2014-4}
\end{figure}

We have also analyzed what happens if we replace the box distribution of $W$ by other distributions, and convinced ourselves that the dependence on the details of the distribution of $W$ is very weak, as the distribution of the finite time Lyapunov exponents is always Gaussian. In Fig.~\ref{figManRob2014-4}(a) we show the histogram of the finite time Lyapunov exponents for $n=100$ with the Gaussian distribution of $W$ with zero mean and standard deviation equal to one.

One might expect that things will be changed drastically if the distribution of $W$ is different, with diverging variance. In Fig.~\ref{figManRob2014-4}(b) we show the result for the Cauchy-Lorentz distribution of $W$ defined as follows
\be  \label{Cauchy-Lorentz}
P(W) = \frac{1}{\pi}\frac{b}{W^2+b^2},
\ee
where $b$ is the halfwidth at the half maximum, and we have chosen $b=1$.
We have taken the values inside the cut-off interval $[-2,+2]$ and $n=100$, and then the same thing for the interval $[-100,+100]$ and $n=100$ in Fig.~\ref{figManRob2014-4}(c). We clearly see that the distribution is always Gaussian.

\section{Numerical study of the decay of the variance of the
distribution of the finite time Lyapunov exponents} \label{sec6}

\begin{figure}
\center
\includegraphics[width=7.5cm]{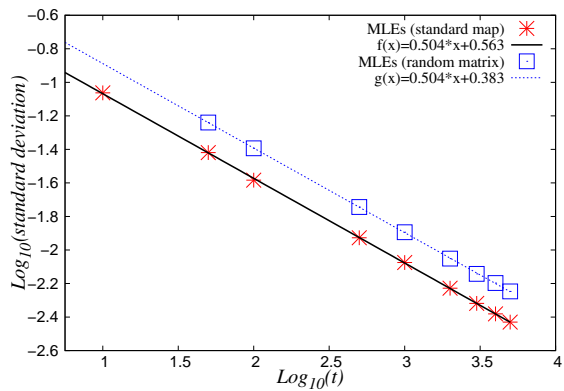}
\caption{[Color online] The standard deviation of the positive finite time
Lyapunov exponents for the  standard map (stars) and  for the product of random transfer matrices with the box distribution of $W$ (empty boxes), as a function of time in $\log-\log$ presentation, and their best fits. The slope is exactly -1/2.}
\label{figManRob2014-8}
\end{figure}

Finally, in this section we present numerical evidence for the theoretical expectation \cite{Ott1993} that the finite time Lyapunov exponents have approximately Gaussian distribution whose variance decreases inversely with time $t$ (the number of iterations in the case of the standard map; Sec.~\ref{sec4}) and $n$ in the case of the product of random matrices in the context of the unimodular transfer matrices of the tight-binding approximation to describe the Anderson localization, expounded in section \ref{sec5}. Indeed, the evidence is overwhelming, as shown in Fig.~\ref{figManRob2014-8}, where we plot the standard deviation as a function of time in log-log plot, showing that it decays inversely with the square root of time.

In the context of our Izrailev model the dimension $N$ of the matrix plays the role of time. The width of the diagonal band is equal to $2k$. Shepelyansky reduces the problem of the localization length to the determination of the smallest positive Lyapunov exponent (its inverse is the localization length)
of the underlying finite dimensional Hamilton system with dimension $2k$. Then, the finite time Lyapunov exponent should have some almost Gaussian distribution, whose mean tends to the asymptotic Lyapunov exponent with $N\rightarrow \infty$ and the variance should decrease to zero as $1/N$.

If this picture were exact, then the mean localization length as a function of $N$ should converge to the asymptotic value, which we do observe in Table~ \ref{tabManRob2014-2} of Sec.~\ref{sec3}, while the variance does {\em not}
decay to zero, but rather remains constant, independent of $N$. From this we conclude that even in the limit $N\rightarrow \infty$ the localization length
has a certain distribution with nonvanishing variance, or more precisely, its inverse (the slope $\sigma$) has an almost Gaussian distribution with nonvanishing variance. We believe that this is the cause of the strong fluctuations observed for example in Fig.~\ref{figManRob2014-7}(a) of Sec.~\ref{sec3}. The same observation applies to the scaling laws of the Figs.~9 and 10 of our previous paper \cite{ManRob2013}.

\section{Summary} \label{sec7}

The main conclusion of this paper is the empirical fact based on our numerical computations of the eigenfunctions of the $N$-dimensional Izrailev model, that the localization length has a distribution with nonvanishing variance not
only for finite $N$, but even in the limit $N\rightarrow \infty$. This is the reason, we believe, for the strong fluctuations in the scaling laws which involve the empirical localization measures and the theoretical semiclassical value of the localization length. In the Shepelyansky picture \cite{She1986}
this might seem to be a contradiction, but the resolution of the puzzle is that in the limit of large $N$ the finite dimensional Hamilton system extracted from the Floquet propagator of the quantum kicked rotator is not good enough, and therefore the matrix elements outside the main diagonal band of width $2k$ play a role, making the Hamilton system effectively infinite dimensional, with infinitely many Lyapunov exponents. This finding is a challenge for the
improved semiclassical theory of the localization length, to derive and explain the discovered distribution function. On the other hand, the simple model of the Anderson localization based on the tight-binding approximation, with only the nearest neighbor interactions,  described by the product of $2\times2$ unimodular matrices, has a finite dimension, as the transfer matrices are exactly two-dimensional, and therefore the variance vanishes in the limit of large dimensions as $1/n$. The same conclusion applies to such a model with a finite number of interacting neighbors. Indeed, according to the references \cite{Kottos1996,Kottos1999} the variance of $\sigma$ should vanish as
$Var(\sigma) \propto 1/(Nk^2)$, but our work shows that in the quantum kicked rotator this is not observed: the variance does not depend on $N$, and decays with $k$ faster than $1/k^2$, namely as $1/k^4$. Thus, here we found some important differences between the dynamical localization in the quantum kicked rotator and the Anderson tight-binding model of localization, and the Shepelyansky picture, which rest upon the banded matrix models with finite bandwidth.

To summarize: We do not have yet a theory to
describe this behavior, namely the theory of the distribution of
the localization length, including the variance, rather than just 
its average value, as explained in the paper, but only the clear 
understanding of what is the reason for this behavior: 
The fact that the banded matrix model for the QKR is not good enough, 
one has to take into account also the (small but many) matrix elements 
outside the main diagonal band, and therefore the Shepelyansky 
picture and approximation breaks down, meaning that the finite dimensional 
Hamiltonian system cannot capture the correct behaviour of the QKR.
Thus, the problem is open for the future work.

\section*{Acknowledgements}

This work was supported by the Slovenian Research Agency (ARRS).

\bibliography{ManRob2014R}

\providecommand{\noopsort}[1]{}\providecommand{\singleletter}[1]{#1}%
\begin{thebibliography}{48}%
\makeatletter
\providecommand \@ifxundefined [1]{%
 \@ifx{#1\undefined}
}%
\providecommand \@ifnum [1]{%
 \ifnum #1\expandafter \@firstoftwo
 \else \expandafter \@secondoftwo
 \fi
}%
\providecommand \@ifx [1]{%
 \ifx #1\expandafter \@firstoftwo
 \else \expandafter \@secondoftwo
 \fi
}%
\providecommand \natexlab [1]{#1}%
\providecommand \enquote  [1]{``#1''}%
\providecommand \bibnamefont  [1]{#1}%
\providecommand \bibfnamefont [1]{#1}%
\providecommand \citenamefont [1]{#1}%
\providecommand \href@noop [0]{\@secondoftwo}%
\providecommand \href [0]{\begingroup \@sanitize@url \@href}%
\providecommand \@href[1]{\@@startlink{#1}\@@href}%
\providecommand \@@href[1]{\endgroup#1\@@endlink}%
\providecommand \@sanitize@url [0]{\catcode `\\12\catcode `\$12\catcode
  `\&12\catcode `\#12\catcode `\^12\catcode `\_12\catcode `\%12\relax}%
\providecommand \@@startlink[1]{}%
\providecommand \@@endlink[0]{}%
\providecommand \url  [0]{\begingroup\@sanitize@url \@url }%
\providecommand \@url [1]{\endgroup\@href {#1}{\urlprefix }}%
\providecommand \urlprefix  [0]{URL }%
\providecommand \Eprint [0]{\href }%
\providecommand \doibase [0]{http://dx.doi.org/}%
\providecommand \selectlanguage [0]{\@gobble}%
\providecommand \bibinfo  [0]{\@secondoftwo}%
\providecommand \bibfield  [0]{\@secondoftwo}%
\providecommand \translation [1]{[#1]}%
\providecommand \BibitemOpen [0]{}%
\providecommand \bibitemStop [0]{}%
\providecommand \bibitemNoStop [0]{.\EOS\space}%
\providecommand \EOS [0]{\spacefactor3000\relax}%
\providecommand \BibitemShut  [1]{\csname bibitem#1\endcsname}%
\let\auto@bib@innerbib\@empty
\bibitem [{\citenamefont {St\"ockmann}(1999)}]{Stoe}%
  \BibitemOpen
  \bibfield  {author} {\bibinfo {author} {\bibfnamefont {H.~J.}\ \bibnamefont
  {St\"ockmann}},\ }\href@noop {} {\emph {\bibinfo {title} {Quantum Chaos - An
  Introduction}}}\ (\bibinfo  {publisher} {Cambridge: Cambridge University
  Press},\ \bibinfo {year} {1999})\BibitemShut {NoStop}%
\bibitem [{\citenamefont {Haake}(2001)}]{Haake}%
  \BibitemOpen
  \bibfield  {author} {\bibinfo {author} {\bibfnamefont {F.}~\bibnamefont
  {Haake}},\ }\href@noop {} {\emph {\bibinfo {title} {Quantum Signatures of
  Chaos}}}\ (\bibinfo  {publisher} {Berlin: Springer},\ \bibinfo {year}
  {2001})\BibitemShut {NoStop}%
\bibitem [{\citenamefont {Izrailev}(1986)}]{Izr1986}%
  \BibitemOpen
  \bibfield  {author} {\bibinfo {author} {\bibfnamefont {F.~M.}\ \bibnamefont
  {Izrailev}},\ }\href@noop {} {\bibfield  {journal} {\bibinfo  {journal}
  {Phys. Rev. Lett.}\ }\textbf {\bibinfo {volume} {56}},\ \bibinfo {pages}
  {541} (\bibinfo {year} {1986})}\BibitemShut {NoStop}%
\bibitem [{\citenamefont {Izrailev}(1987)}]{Izr1987}%
  \BibitemOpen
  \bibfield  {author} {\bibinfo {author} {\bibfnamefont {F.~M.}\ \bibnamefont
  {Izrailev}},\ }\href@noop {} {\bibfield  {journal} {\bibinfo  {journal}
  {Phys. Lett. A}\ }\textbf {\bibinfo {volume} {125}},\ \bibinfo {pages} {250}
  (\bibinfo {year} {1987})}\BibitemShut {NoStop}%
\bibitem [{\citenamefont {Izrailev}(1989)}]{Izr1989}%
  \BibitemOpen
  \bibfield  {author} {\bibinfo {author} {\bibfnamefont {F.~M.}\ \bibnamefont
  {Izrailev}},\ }\href@noop {} {\bibfield  {journal} {\bibinfo  {journal} {J.
  Phys. A: Math. Gen.}\ }\textbf {\bibinfo {volume} {22}},\ \bibinfo {pages}
  {865} (\bibinfo {year} {1989})}\BibitemShut {NoStop}%
\bibitem [{\citenamefont {Izrailev}(1990)}]{Izr1990}%
  \BibitemOpen
  \bibfield  {author} {\bibinfo {author} {\bibfnamefont {F.~M.}\ \bibnamefont
  {Izrailev}},\ }\href@noop {} {\bibfield  {journal} {\bibinfo  {journal}
  {Phys. Rep.}\ }\textbf {\bibinfo {volume} {196}},\ \bibinfo {pages} {299}
  (\bibinfo {year} {1990})}\BibitemShut {NoStop}%
\bibitem [{\citenamefont {Manos}\ and\ \citenamefont
  {Robnik}(2014)}]{ManRob2014}%
  \BibitemOpen
  \bibfield  {author} {\bibinfo {author} {\bibfnamefont {T.}~\bibnamefont
  {Manos}}\ and\ \bibinfo {author} {\bibfnamefont {M.}~\bibnamefont {Robnik}},\
  }\href@noop {} {\bibfield  {journal} {\bibinfo  {journal} {Phys. Rev. E}\
  }\textbf {\bibinfo {volume} {89}},\ \bibinfo {pages} {022905} (\bibinfo
  {year} {2014})}\BibitemShut {NoStop}%
\bibitem [{\citenamefont {Shepelyansky}(1986)}]{She1986}%
  \BibitemOpen
  \bibfield  {author} {\bibinfo {author} {\bibfnamefont {D.~L.}\ \bibnamefont
  {Shepelyansky}},\ }\href@noop {} {\bibfield  {journal} {\bibinfo  {journal}
  {Phys. Rev. Lett.}\ }\textbf {\bibinfo {volume} {56}},\ \bibinfo {pages}
  {677} (\bibinfo {year} {1986})}\BibitemShut {NoStop}%
\bibitem [{\citenamefont {Manos}\ and\ \citenamefont
  {Robnik}(2013)}]{ManRob2013}%
  \BibitemOpen
  \bibfield  {author} {\bibinfo {author} {\bibfnamefont {T.}~\bibnamefont
  {Manos}}\ and\ \bibinfo {author} {\bibfnamefont {M.}~\bibnamefont {Robnik}},\
  }\href@noop {} {\bibfield  {journal} {\bibinfo  {journal} {Phys. Rev. E}\
  }\textbf {\bibinfo {volume} {87}},\ \bibinfo {pages} {062905} (\bibinfo
  {year} {2013})}\BibitemShut {NoStop}%
\bibitem [{\citenamefont {Robnik}(1998)}]{Rob1998}%
  \BibitemOpen
  \bibfield  {author} {\bibinfo {author} {\bibfnamefont {M.}~\bibnamefont
  {Robnik}},\ }\href@noop {} {\bibfield  {journal} {\bibinfo  {journal} {Nonl.
  Phen. in Compl. Syst. (Minsk)}\ }\textbf {\bibinfo {volume} {1}},\ \bibinfo
  {pages} {1} (\bibinfo {year} {1998})}\BibitemShut {NoStop}%
\bibitem [{\citenamefont {Mehta}(1991)}]{Mehta}%
  \BibitemOpen
  \bibfield  {author} {\bibinfo {author} {\bibfnamefont {M.~L.}\ \bibnamefont
  {Mehta}},\ }\href@noop {} {\emph {\bibinfo {title} {Random Matrices}}}\
  (\bibinfo  {publisher} {Boston: Academic Press},\ \bibinfo {year}
  {1991})\BibitemShut {NoStop}%
\bibitem [{\citenamefont {Guhr}\ \emph {et~al.}(1998)\citenamefont {Guhr},
  \citenamefont {M\"uller-Groeling},\ and\ \citenamefont
  {Weidenm\"uller}}]{GMW}%
  \BibitemOpen
  \bibfield  {author} {\bibinfo {author} {\bibfnamefont {T.}~\bibnamefont
  {Guhr}}, \bibinfo {author} {\bibfnamefont {A.}~\bibnamefont
  {M\"uller-Groeling}}, \ and\ \bibinfo {author} {\bibfnamefont
  {H.}~\bibnamefont {Weidenm\"uller}},\ }\href@noop {} {\bibfield  {journal}
  {\bibinfo  {journal} {Phys. Rep.}\ }\textbf {\bibinfo {volume} {299}},\
  \bibinfo {pages} {4} (\bibinfo {year} {1998})}\BibitemShut {NoStop}%
\bibitem [{\citenamefont {Bohigas}\ \emph {et~al.}(1984)\citenamefont
  {Bohigas}, \citenamefont {Giannoni},\ and\ \citenamefont {Schmit}}]{BGS}%
  \BibitemOpen
  \bibfield  {author} {\bibinfo {author} {\bibfnamefont {O.}~\bibnamefont
  {Bohigas}}, \bibinfo {author} {\bibfnamefont {M.~J.}\ \bibnamefont
  {Giannoni}}, \ and\ \bibinfo {author} {\bibfnamefont {C.}~\bibnamefont
  {Schmit}},\ }\href@noop {} {\bibfield  {journal} {\bibinfo  {journal} {Phys.
  Rev. Lett.}\ }\textbf {\bibinfo {volume} {52}},\ \bibinfo {pages} {1}
  (\bibinfo {year} {1984})}\BibitemShut {NoStop}%
\bibitem [{\citenamefont {Casati}\ \emph {et~al.}(1980)\citenamefont {Casati},
  \citenamefont {Valz-Gris},\ and\ \citenamefont {Guarneri}}]{Cas}%
  \BibitemOpen
  \bibfield  {author} {\bibinfo {author} {\bibfnamefont {G.}~\bibnamefont
  {Casati}}, \bibinfo {author} {\bibfnamefont {F.}~\bibnamefont {Valz-Gris}}, \
  and\ \bibinfo {author} {\bibfnamefont {I.}~\bibnamefont {Guarneri}},\
  }\href@noop {} {\bibfield  {journal} {\bibinfo  {journal} {Lett. Nuovo
  Cimento}\ }\textbf {\bibinfo {volume} {28}},\ \bibinfo {pages} {279}
  (\bibinfo {year} {1980})}\BibitemShut {NoStop}%
\bibitem [{\citenamefont {Robnik}\ and\ \citenamefont {Berry}(1986)}]{RB1986}%
  \BibitemOpen
  \bibfield  {author} {\bibinfo {author} {\bibfnamefont {M.}~\bibnamefont
  {Robnik}}\ and\ \bibinfo {author} {\bibfnamefont {M.~V.}\ \bibnamefont
  {Berry}},\ }\href@noop {} {\bibfield  {journal} {\bibinfo  {journal} {J.
  Phys. A: Math. Gen.}\ }\textbf {\bibinfo {volume} {19}},\ \bibinfo {pages}
  {669} (\bibinfo {year} {1986})}\BibitemShut {NoStop}%
\bibitem [{\citenamefont {Robnik}(1986)}]{Rob1986}%
  \BibitemOpen
  \bibfield  {author} {\bibinfo {author} {\bibfnamefont {M.}~\bibnamefont
  {Robnik}},\ }\href@noop {} {\bibfield  {journal} {\bibinfo  {journal} {Lect.
  Notes Phys.}\ }\textbf {\bibinfo {volume} {263}},\ \bibinfo {pages} {120}
  (\bibinfo {year} {1986})}\BibitemShut {NoStop}%
\bibitem [{\citenamefont {Berry}(1985)}]{Berry1985}%
  \BibitemOpen
  \bibfield  {author} {\bibinfo {author} {\bibfnamefont {M.~V.}\ \bibnamefont
  {Berry}},\ }\href@noop {} {\bibfield  {journal} {\bibinfo  {journal} {Proc.
  Roy. Soc. Lond. A}\ }\textbf {\bibinfo {volume} {400}},\ \bibinfo {pages}
  {229} (\bibinfo {year} {1985})}\BibitemShut {NoStop}%
\bibitem [{\citenamefont {Sieber}\ and\ \citenamefont
  {Richter}(2001)}]{Sieber}%
  \BibitemOpen
  \bibfield  {author} {\bibinfo {author} {\bibfnamefont {M.}~\bibnamefont
  {Sieber}}\ and\ \bibinfo {author} {\bibfnamefont {K.}~\bibnamefont
  {Richter}},\ }\href@noop {} {\bibfield  {journal} {\bibinfo  {journal} {Phys.
  Scr.}\ }\textbf {\bibinfo {volume} {T90}},\ \bibinfo {pages} {128} (\bibinfo
  {year} {2001})}\BibitemShut {NoStop}%
\bibitem [{\citenamefont {M\"uller}\ \emph {et~al.}(2004)\citenamefont
  {M\"uller}, \citenamefont {Heusler}, \citenamefont {Braun}, \citenamefont
  {Haake},\ and\ \citenamefont {Altland}}]{Mueller1}%
  \BibitemOpen
  \bibfield  {author} {\bibinfo {author} {\bibfnamefont {S.}~\bibnamefont
  {M\"uller}}, \bibinfo {author} {\bibfnamefont {S.}~\bibnamefont {Heusler}},
  \bibinfo {author} {\bibfnamefont {P.}~\bibnamefont {Braun}}, \bibinfo
  {author} {\bibfnamefont {F.}~\bibnamefont {Haake}}, \ and\ \bibinfo {author}
  {\bibfnamefont {A.}~\bibnamefont {Altland}},\ }\href@noop {} {\bibfield
  {journal} {\bibinfo  {journal} {Phys. Rev. Lett.}\ }\textbf {\bibinfo
  {volume} {93}},\ \bibinfo {pages} {014103} (\bibinfo {year}
  {2004})}\BibitemShut {NoStop}%
\bibitem [{\citenamefont {Heusler}\ \emph {et~al.}(2004)\citenamefont
  {Heusler}, \citenamefont {M\"uller}, \citenamefont {Braun},\ and\
  \citenamefont {Haake}}]{Mueller2}%
  \BibitemOpen
  \bibfield  {author} {\bibinfo {author} {\bibfnamefont {S.}~\bibnamefont
  {Heusler}}, \bibinfo {author} {\bibfnamefont {S.}~\bibnamefont {M\"uller}},
  \bibinfo {author} {\bibfnamefont {P.}~\bibnamefont {Braun}}, \ and\ \bibinfo
  {author} {\bibfnamefont {F.}~\bibnamefont {Haake}},\ }\href@noop {}
  {\bibfield  {journal} {\bibinfo  {journal} {J. Phys.A: Math. Gen.}\ }\textbf
  {\bibinfo {volume} {37}},\ \bibinfo {pages} {L31} (\bibinfo {year}
  {2004})}\BibitemShut {NoStop}%
\bibitem [{\citenamefont {M\"uller}\ \emph {et~al.}(2005)\citenamefont
  {M\"uller}, \citenamefont {Heusler}, \citenamefont {Braun}, \citenamefont
  {Haake},\ and\ \citenamefont {Altland}}]{Mueller3}%
  \BibitemOpen
  \bibfield  {author} {\bibinfo {author} {\bibfnamefont {S.}~\bibnamefont
  {M\"uller}}, \bibinfo {author} {\bibfnamefont {S.}~\bibnamefont {Heusler}},
  \bibinfo {author} {\bibfnamefont {P.}~\bibnamefont {Braun}}, \bibinfo
  {author} {\bibfnamefont {F.}~\bibnamefont {Haake}}, \ and\ \bibinfo {author}
  {\bibfnamefont {A.}~\bibnamefont {Altland}},\ }\href@noop {} {\bibfield
  {journal} {\bibinfo  {journal} {Phys. Rev. E}\ }\textbf {\bibinfo {volume}
  {72}},\ \bibinfo {pages} {046207} (\bibinfo {year} {2005})}\BibitemShut
  {NoStop}%
\bibitem [{\citenamefont {M\"uller}\ \emph {et~al.}(2009)\citenamefont
  {M\"uller}, \citenamefont {Heusler}, \citenamefont {Altland}, \citenamefont
  {Braun},\ and\ \citenamefont {Haake}}]{Mueller4}%
  \BibitemOpen
  \bibfield  {author} {\bibinfo {author} {\bibfnamefont {S.}~\bibnamefont
  {M\"uller}}, \bibinfo {author} {\bibfnamefont {S.}~\bibnamefont {Heusler}},
  \bibinfo {author} {\bibfnamefont {A.}~\bibnamefont {Altland}}, \bibinfo
  {author} {\bibfnamefont {P.}~\bibnamefont {Braun}}, \ and\ \bibinfo {author}
  {\bibfnamefont {F.}~\bibnamefont {Haake}},\ }\href@noop {} {\bibfield
  {journal} {\bibinfo  {journal} {New J. of Phys.}\ }\textbf {\bibinfo {volume}
  {11}},\ \bibinfo {pages} {103025} (\bibinfo {year} {2009})}\BibitemShut
  {NoStop}%
\bibitem [{\citenamefont {Prosen}()}]{Pro2000}%
  \BibitemOpen
  \bibfield  {author} {\bibinfo {author} {\bibfnamefont {T.}~\bibnamefont
  {Prosen}},\ }\href@noop {} {}\bibinfo {howpublished} {\textit{in Proceedings
  of the International School of Physics ``Enrico Fermi'', Course CXLIII}},\
  \bibinfo {note} {edited by G. Casati and I. Guarneri and U. Smilyanski
  (Amsterdam: IOS Press, 2000) p.~473}\BibitemShut {NoStop}%
\bibitem [{\citenamefont {Brody}(1973)}]{Bro1973}%
  \BibitemOpen
  \bibfield  {author} {\bibinfo {author} {\bibfnamefont {T.~A.}\ \bibnamefont
  {Brody}},\ }\href@noop {} {\bibfield  {journal} {\bibinfo  {journal} {Lett.
  Nuovo Cimento}\ }\textbf {\bibinfo {volume} {7}},\ \bibinfo {pages} {482}
  (\bibinfo {year} {1973})}\BibitemShut {NoStop}%
\bibitem [{\citenamefont {Brody}\ \emph {et~al.}(1981)\citenamefont {Brody},
  \citenamefont {Flores}, \citenamefont {French}, \citenamefont {Mello},
  \citenamefont {Pandey},\ and\ \citenamefont {Wong}}]{Bro1981}%
  \BibitemOpen
  \bibfield  {author} {\bibinfo {author} {\bibfnamefont {T.~A.}\ \bibnamefont
  {Brody}}, \bibinfo {author} {\bibfnamefont {J.}~\bibnamefont {Flores}},
  \bibinfo {author} {\bibfnamefont {J.~B.}\ \bibnamefont {French}}, \bibinfo
  {author} {\bibfnamefont {P.~A.}\ \bibnamefont {Mello}}, \bibinfo {author}
  {\bibfnamefont {A.}~\bibnamefont {Pandey}}, \ and\ \bibinfo {author}
  {\bibfnamefont {S.~S.~M.}\ \bibnamefont {Wong}},\ }\href@noop {} {\bibfield
  {journal} {\bibinfo  {journal} {Rev. Mod. Phys.}\ }\textbf {\bibinfo {volume}
  {53}},\ \bibinfo {pages} {385} (\bibinfo {year} {1981})}\BibitemShut
  {NoStop}%
\bibitem [{\citenamefont {Batisti\'c}\ and\ \citenamefont
  {Robnik}(2010)}]{BatRob2010}%
  \BibitemOpen
  \bibfield  {author} {\bibinfo {author} {\bibfnamefont {B.}~\bibnamefont
  {Batisti\'c}}\ and\ \bibinfo {author} {\bibfnamefont {M.}~\bibnamefont
  {Robnik}},\ }\href@noop {} {\bibfield  {journal} {\bibinfo  {journal} {J.
  Phys. A: Math. Gen.}\ }\textbf {\bibinfo {volume} {43}},\ \bibinfo {pages}
  {215101} (\bibinfo {year} {2010})}\BibitemShut {NoStop}%
\bibitem [{\citenamefont {Batisti\'c}\ \emph {et~al.}(2013)\citenamefont
  {Batisti\'c}, \citenamefont {Manos},\ and\ \citenamefont {Robnik}}]{BMR2013}%
  \BibitemOpen
  \bibfield  {author} {\bibinfo {author} {\bibfnamefont {B.}~\bibnamefont
  {Batisti\'c}}, \bibinfo {author} {\bibfnamefont {T.}~\bibnamefont {Manos}}, \
  and\ \bibinfo {author} {\bibfnamefont {M.}~\bibnamefont {Robnik}},\
  }\href@noop {} {\bibfield  {journal} {\bibinfo  {journal} {Europhys. Lett.}\
  }\textbf {\bibinfo {volume} {102}},\ \bibinfo {pages} {50008} (\bibinfo
  {year} {2013})}\BibitemShut {NoStop}%
\bibitem [{\citenamefont {Batisti\'c}\ and\ \citenamefont
  {Robnik}(2013{\natexlab{a}})}]{BatRob2013}%
  \BibitemOpen
  \bibfield  {author} {\bibinfo {author} {\bibfnamefont {B.}~\bibnamefont
  {Batisti\'c}}\ and\ \bibinfo {author} {\bibfnamefont {M.}~\bibnamefont
  {Robnik}},\ }\href@noop {} {\bibfield  {journal} {\bibinfo  {journal} {J.
  Phys. A: Math. Theor.}\ }\textbf {\bibinfo {volume} {46}},\ \bibinfo {pages}
  {315102} (\bibinfo {year} {2013}{\natexlab{a}})}\BibitemShut {NoStop}%
\bibitem [{\citenamefont {Batisti\'c}\ and\ \citenamefont
  {Robnik}(2013{\natexlab{b}})}]{BatRob2013A}%
  \BibitemOpen
  \bibfield  {author} {\bibinfo {author} {\bibfnamefont {B.}~\bibnamefont
  {Batisti\'c}}\ and\ \bibinfo {author} {\bibfnamefont {M.}~\bibnamefont
  {Robnik}},\ }\href@noop {} {\bibfield  {journal} {\bibinfo  {journal} {Phys.
  Rev. E}\ }\textbf {\bibinfo {volume} {88}},\ \bibinfo {pages} {052913}
  (\bibinfo {year} {2013}{\natexlab{b}})}\BibitemShut {NoStop}%
\bibitem [{\citenamefont {Casati}\ \emph {et~al.}(1979)\citenamefont {Casati},
  \citenamefont {Chirikov}, \citenamefont {Ford},\ and\ \citenamefont
  {Izrailev}}]{CCFI79}%
  \BibitemOpen
  \bibfield  {author} {\bibinfo {author} {\bibfnamefont {G.}~\bibnamefont
  {Casati}}, \bibinfo {author} {\bibfnamefont {B.}~\bibnamefont {Chirikov}},
  \bibinfo {author} {\bibfnamefont {J.}~\bibnamefont {Ford}}, \ and\ \bibinfo
  {author} {\bibfnamefont {F.~M.}\ \bibnamefont {Izrailev}},\ }\href@noop {}
  {\bibfield  {journal} {\bibinfo  {journal} {Lect. Notes Phys.}\ }\textbf
  {\bibinfo {volume} {93}},\ \bibinfo {pages} {334} (\bibinfo {year}
  {1979})}\BibitemShut {NoStop}%
\bibitem [{\citenamefont {Taylor}()}]{T69}%
  \BibitemOpen
  \bibfield  {author} {\bibinfo {author} {\bibfnamefont {J.~B.}\ \bibnamefont
  {Taylor}},\ }\href@noop {} {}\bibinfo {note} {Culham Laboratory Progress
  Report, CLM-PR-12 (1969)}\BibitemShut {NoStop}%
\bibitem [{\citenamefont {Froeschl\'e}(1970)}]{F72}%
  \BibitemOpen
  \bibfield  {author} {\bibinfo {author} {\bibfnamefont {C.}~\bibnamefont
  {Froeschl\'e}},\ }\href@noop {} {\bibfield  {journal} {\bibinfo  {journal}
  {Astron. Astrophys.}\ }\textbf {\bibinfo {volume} {9}},\ \bibinfo {pages}
  {15} (\bibinfo {year} {1970})}\BibitemShut {NoStop}%
\bibitem [{\citenamefont {Chirikov}(1979)}]{C79}%
  \BibitemOpen
  \bibfield  {author} {\bibinfo {author} {\bibfnamefont {B.}~\bibnamefont
  {Chirikov}},\ }\href@noop {} {\bibfield  {journal} {\bibinfo  {journal}
  {Phys. Rep.}\ }\textbf {\bibinfo {volume} {52}},\ \bibinfo {pages} {263}
  (\bibinfo {year} {1979})}\BibitemShut {NoStop}%
\bibitem [{\citenamefont {Zeldovich}(1966)}]{Zel1966}%
  \BibitemOpen
  \bibfield  {author} {\bibinfo {author} {\bibfnamefont {Y.~B.}\ \bibnamefont
  {Zeldovich}},\ }\href@noop {} {\bibfield  {journal} {\bibinfo  {journal}
  {Eksp. Teor. Fiz.}\ }\textbf {\bibinfo {volume} {51}},\ \bibinfo {pages}
  {1942} (\bibinfo {year} {1966})}\BibitemShut {NoStop}%
\bibitem [{\citenamefont {Izrailev}\ and\ \citenamefont
  {Shepelyansky}(1979{\natexlab{a}})}]{IS1979a}%
  \BibitemOpen
  \bibfield  {author} {\bibinfo {author} {\bibfnamefont {F.~M.}\ \bibnamefont
  {Izrailev}}\ and\ \bibinfo {author} {\bibfnamefont {D.~L.}\ \bibnamefont
  {Shepelyansky}},\ }\href@noop {} {\bibfield  {journal} {\bibinfo  {journal}
  {Dokl. Akad. Nauk SSSR}\ }\textbf {\bibinfo {volume} {249}},\ \bibinfo
  {pages} {1103} (\bibinfo {year} {1979}{\natexlab{a}})}\BibitemShut {NoStop}%
\bibitem [{\citenamefont {Izrailev}\ and\ \citenamefont
  {Shepelyansky}(1979{\natexlab{b}})}]{IS1979b}%
  \BibitemOpen
  \bibfield  {author} {\bibinfo {author} {\bibfnamefont {F.~M.}\ \bibnamefont
  {Izrailev}}\ and\ \bibinfo {author} {\bibfnamefont {D.~L.}\ \bibnamefont
  {Shepelyansky}},\ }\href@noop {} {\bibfield  {journal} {\bibinfo  {journal}
  {Sov. Phys. Dokl.}\ }\textbf {\bibinfo {volume} {24}},\ \bibinfo {pages}
  {996} (\bibinfo {year} {1979}{\natexlab{b}})}\BibitemShut {NoStop}%
\bibitem [{\citenamefont {Izrailev}\ and\ \citenamefont
  {Shepelyansky}(1980{\natexlab{a}})}]{IS1980a}%
  \BibitemOpen
  \bibfield  {author} {\bibinfo {author} {\bibfnamefont {F.~M.}\ \bibnamefont
  {Izrailev}}\ and\ \bibinfo {author} {\bibfnamefont {D.~L.}\ \bibnamefont
  {Shepelyansky}},\ }\href@noop {} {\bibfield  {journal} {\bibinfo  {journal}
  {Teor. Mat. Fiz.}\ }\textbf {\bibinfo {volume} {43}},\ \bibinfo {pages} {417}
  (\bibinfo {year} {1980}{\natexlab{a}})}\BibitemShut {NoStop}%
\bibitem [{\citenamefont {Izrailev}\ and\ \citenamefont
  {Shepelyansky}(1980{\natexlab{b}})}]{IS1980b}%
  \BibitemOpen
  \bibfield  {author} {\bibinfo {author} {\bibfnamefont {F.~M.}\ \bibnamefont
  {Izrailev}}\ and\ \bibinfo {author} {\bibfnamefont {D.~L.}\ \bibnamefont
  {Shepelyansky}},\ }\href@noop {} {\bibfield  {journal} {\bibinfo  {journal}
  {Theor. Math. Phys.}\ }\textbf {\bibinfo {volume} {43}},\ \bibinfo {pages}
  {553} (\bibinfo {year} {1980}{\natexlab{b}})}\BibitemShut {NoStop}%
\bibitem [{\citenamefont {Chirikov}\ \emph {et~al.}(1981)\citenamefont
  {Chirikov}, \citenamefont {Izrailev},\ and\ \citenamefont
  {Shepelyansky}}]{CIS1981}%
  \BibitemOpen
  \bibfield  {author} {\bibinfo {author} {\bibfnamefont {B.~V.}\ \bibnamefont
  {Chirikov}}, \bibinfo {author} {\bibfnamefont {F.~M.}\ \bibnamefont
  {Izrailev}}, \ and\ \bibinfo {author} {\bibfnamefont {D.~L.}\ \bibnamefont
  {Shepelyansky}},\ }\href@noop {} {\bibfield  {journal} {\bibinfo  {journal}
  {Sov. Sci. Revv. C 2}\ }\textbf {\bibinfo {volume} {2}},\ \bibinfo {pages}
  {209} (\bibinfo {year} {1981})}\BibitemShut {NoStop}%
\bibitem [{\citenamefont {Izrailev}(1988)}]{Izr1988}%
  \BibitemOpen
  \bibfield  {author} {\bibinfo {author} {\bibfnamefont {F.~M.}\ \bibnamefont
  {Izrailev}},\ }\href@noop {} {\bibfield  {journal} {\bibinfo  {journal}
  {Phys. Lett. A}\ }\textbf {\bibinfo {volume} {134}},\ \bibinfo {pages} {13}
  (\bibinfo {year} {1988})}\BibitemShut {NoStop}%
\bibitem [{\citenamefont {Lichtenberg}\ and\ \citenamefont
  {Lieberman}(1992)}]{LL1992}%
  \BibitemOpen
  \bibfield  {author} {\bibinfo {author} {\bibfnamefont {A.~J.}\ \bibnamefont
  {Lichtenberg}}\ and\ \bibinfo {author} {\bibfnamefont {M.~A.}\ \bibnamefont
  {Lieberman}},\ }\href@noop {} {\emph {\bibinfo {title} {Regular and Chaotic
  Dynamics}}}\ (\bibinfo  {publisher} {New York: Springer Verlag},\ \bibinfo
  {year} {1992})\BibitemShut {NoStop}%
\bibitem [{\citenamefont {Kottos}\ \emph {et~al.}(1996)\citenamefont {Kottos},
  \citenamefont {Politi}, \citenamefont {Izrailev},\ and\ \citenamefont
  {Ruffo}}]{Kottos1996}%
  \BibitemOpen
  \bibfield  {author} {\bibinfo {author} {\bibfnamefont {T.}~\bibnamefont
  {Kottos}}, \bibinfo {author} {\bibfnamefont {A.}~\bibnamefont {Politi}},
  \bibinfo {author} {\bibfnamefont {F.}~\bibnamefont {Izrailev}}, \ and\
  \bibinfo {author} {\bibfnamefont {S.}~\bibnamefont {Ruffo}},\ }\href@noop {}
  {\bibfield  {journal} {\bibinfo  {journal} {Phys. Rev. E}\ }\textbf {\bibinfo
  {volume} {53}},\ \bibinfo {pages} {R5553} (\bibinfo {year}
  {1996})}\BibitemShut {NoStop}%
\bibitem [{\citenamefont {Fujisaka}(1983)}]{Fuj1983}%
  \BibitemOpen
  \bibfield  {author} {\bibinfo {author} {\bibfnamefont {H.}~\bibnamefont
  {Fujisaka}},\ }\href@noop {} {\bibfield  {journal} {\bibinfo  {journal}
  {Prog. Theor. Phys.}\ }\textbf {\bibinfo {volume} {70}},\ \bibinfo {pages}
  {1264} (\bibinfo {year} {1983})}\BibitemShut {NoStop}%
\bibitem [{\citenamefont {Ott}(1993)}]{Ott1993}%
  \BibitemOpen
  \bibfield  {author} {\bibinfo {author} {\bibfnamefont {E.}~\bibnamefont
  {Ott}},\ }\href@noop {} {\emph {\bibinfo {title} {Chaos in Dynamical
  Systems}}}\ (\bibinfo  {publisher} {Cambridge University Press},\ \bibinfo
  {year} {1993})\BibitemShut {NoStop}%
\bibitem [{\citenamefont {Fishman}\ \emph {et~al.}(1982)\citenamefont
  {Fishman}, \citenamefont {Grempel},\ and\ \citenamefont {Prange}}]{FGP1982}%
  \BibitemOpen
  \bibfield  {author} {\bibinfo {author} {\bibfnamefont {S.}~\bibnamefont
  {Fishman}}, \bibinfo {author} {\bibfnamefont {D.}~\bibnamefont {Grempel}}, \
  and\ \bibinfo {author} {\bibfnamefont {R.}~\bibnamefont {Prange}},\
  }\href@noop {} {\bibfield  {journal} {\bibinfo  {journal} {Phys. Rev. Lett.}\
  }\textbf {\bibinfo {volume} {49}},\ \bibinfo {pages} {509} (\bibinfo {year}
  {1982})}\BibitemShut {NoStop}%
\bibitem [{\citenamefont {Prange}\ \emph {et~al.}(1984)\citenamefont {Prange},
  \citenamefont {Grempel},\ and\ \citenamefont {Fishman}}]{PGF1984}%
  \BibitemOpen
  \bibfield  {author} {\bibinfo {author} {\bibfnamefont {R.}~\bibnamefont
  {Prange}}, \bibinfo {author} {\bibfnamefont {D.}~\bibnamefont {Grempel}}, \
  and\ \bibinfo {author} {\bibfnamefont {S.}~\bibnamefont {Fishman}},\
  }\href@noop {} {\emph {\bibinfo {title} {Como Conference on Quantum Chaos, G.
  Casati, ed.}}}\ (\bibinfo  {publisher} {Plenum, New York},\ \bibinfo {year}
  {1984})\BibitemShut {NoStop}%
\bibitem [{\citenamefont {Crisanti}\ \emph {et~al.}(1993)\citenamefont
  {Crisanti}, \citenamefont {Paladin},\ and\ \citenamefont
  {Vulpiani}}]{CPV1993}%
  \BibitemOpen
  \bibfield  {author} {\bibinfo {author} {\bibfnamefont {A.}~\bibnamefont
  {Crisanti}}, \bibinfo {author} {\bibfnamefont {G.}~\bibnamefont {Paladin}}, \
  and\ \bibinfo {author} {\bibfnamefont {A.}~\bibnamefont {Vulpiani}},\
  }\href@noop {} {\emph {\bibinfo {title} {Products of Random Matrices in
  Statistical Physics}}}\ (\bibinfo  {publisher} {Springer-Verlag, Berlin
  Heidelberg},\ \bibinfo {year} {1993})\BibitemShut {NoStop}%
\bibitem [{\citenamefont {Kottos}\ \emph {et~al.}(1999)\citenamefont {Kottos},
  \citenamefont {Izrailev},\ and\ \citenamefont {Politi}}]{Kottos1999}%
  \BibitemOpen
  \bibfield  {author} {\bibinfo {author} {\bibfnamefont {T.}~\bibnamefont
  {Kottos}}, \bibinfo {author} {\bibfnamefont {F.}~\bibnamefont {Izrailev}}, \
  and\ \bibinfo {author} {\bibfnamefont {A.}~\bibnamefont {Politi}},\
  }\href@noop {} {\bibfield  {journal} {\bibinfo  {journal} {Physica D}\
  }\textbf {\bibinfo {volume} {131}},\ \bibinfo {pages} {155} (\bibinfo {year}
  {1999})}\BibitemShut {NoStop}%
\end{thebibliography}%

\end{document}